\documentclass{llncs}
\usepackage{rotating}
\usepackage[final]{microtype}
\usepackage{color}
\usepackage{latexsym}
\usepackage{amssymb}
\usepackage{amsmath}
\usepackage{xspace}
\usepackage{multirow}
\usepackage{placeins}
\usepackage{wrapfig}
\usepackage[caption=false, position=top]{subfig}
\usepackage{url}
\usepackage{bussproofs}
\usepackage{hyperref}

\author{Florian Lonsing \and Uwe Egly}

\institute{Research Division of Knowledge Based Systems \\ 
  Institute of Logic and Computation,  
  TU Wien, Vienna, Austria \\ 
  \email{\{florian.lonsing,uwe.egly\}@tuwien.ac.at}
}

\begin{document}

\newcommand{\hqspre}{\textsf{HQSpre}\xspace}
\newcommand{\hqspreg}{\textsf{HQSpre-g}\xspace}
\newcommand{\bloqqer}{\textsf{Bloqqer}\xspace}
\newcommand{\aigsolve}{\textsf{AIGSolve}\xspace}
\newcommand{\ghostqcegar}{\textsf{GhostQ}\xspace}
\newcommand{\ghostqcegarnodef}{\textsf{GhostQ-nd}\xspace}
\newcommand{\qsts}{\textsf{QSTS}\xspace}
\newcommand{\qstsdefnobreaksym}{\textsf{QSTS}\xspace}
\newcommand{\qstsbreaksym}{\textsf{QSTS-dsb}\xspace}
\newcommand{\qstsnodef}{\textsf{QSTS-ndsb}\xspace}
\newcommand{\depqbfbat}{\textsf{DepQBF}\xspace}
\newcommand{\depqbfnobat}{\textsf{DepQBF-n}\xspace}
\newcommand{\caqe}{\textsf{CAQE}\xspace}
\newcommand{\qesto}{\textsf{QESTO}\xspace}
\newcommand{\rareqs}{\textsf{RAReQS}\xspace}
\newcommand{\dynqbf}{\textsf{DynQBF}\xspace}
\newcommand{\heretic}{\textsf{Heretic}\xspace}
\newcommand{\ijtihad}{\textsf{Ijtihad}\xspace}
\newcommand{\quterandom}{\textsf{Qute}\xspace}
\newcommand{\revqfun}{\textsf{Rev-Qfun}\xspace}
\newcommand{\depqbfprefixopt}{\textsf{DepQBF}\xspace}

\newcommand{\instancesnopreall}{S_{16\mid825}\xspace}
\newcommand{\instancespreall}{S_{16\mid825}'\xspace}
\newcommand{\instancesnoprefiltered}{S_{16\mid402}\xspace}
\newcommand{\instancesprefiltered}{S_{16\mid402}'\xspace}
\newcommand{\instancesnoprefilteredHQSpre}{S_{16\mid308}\xspace}
\newcommand{\instancesprefilteredHQSpre}{S_{16\mid308}'\xspace}
\newcommand{\instancesnoprefilteredHQSpreG}{S_{16\mid515}\xspace}
\newcommand{\instancesprefilteredHQSpreG}{S_{16\mid515}'\xspace}
\newcommand{\instancesnopreallNEW}{S_{17\mid523}\xspace}
\newcommand{\instancespreallNEW}{S_{17\mid523}'\xspace}
\newcommand{\instancesnoprefilteredNEW}{S_{17\mid437}\xspace}
\newcommand{\instancesprefilteredNEW}{S_{17\mid437}'\xspace}
\newcommand{\instancesnoprefilteredHQSpreNEW}{S_{17\mid312}\xspace}
\newcommand{\instancesprefilteredHQSpreNEW}{S_{17\mid312}'\xspace}
\newcommand{\instancesnoprefilteredHQSpreGNEW}{S_{17\mid380}\xspace}
\newcommand{\instancesprefilteredHQSpreGNEW}{S_{17\mid380}'\xspace}

\newcommand{\doctitle}{Evaluating QBF Solvers:\\Quantifier Alternations Matter}

\title{\doctitle
\thanks{
Supported by the Austrian Science Fund (FWF) under grant
  S11409-N23. This article will appear in the \textbf{proceedings} of CP 2018,
  LNCS, Springer, 2018.
}}

\maketitle

\begin{abstract}
We present an experimental study of the effects of quantifier alternations on
the evaluation of quantified Boolean formula (QBF) solvers.  The number of
quantifier alternations in a QBF in prenex conjunctive normal form (PCNF) is
directly related to the theoretical hardness of the respective QBF
satisfiability problem in the polynomial hierarchy. 
We show empirically that the performance of solvers based on
different solving paradigms substantially varies depending on 
the numbers of alternations in PCNFs. In related theoretical work, quantifier
alternations have become the focus of understanding the strengths and
weaknesses of various QBF proof systems implemented in solvers.  Our results
motivate the development of methods to evaluate orthogonal solving paradigms
by taking quantifier alternations into account.  This is necessary to showcase
the broad range of existing QBF solving paradigms for practical QBF
applications. Moreover, we highlight the potential of combining different
approaches and QBF proof systems in solvers.
\end{abstract}

%%%%%%%%%%%%%%%%%%%%%%%%%%%%%%%%%%%%%%%%%%%%%%%%%%%%%%%%%%%%%%%%%%%%%%%%%%%%%%%%
%%%%%%%%%%%%%%%%%%%%%%%%%%%%%%%%%%%%%%%%%%%%%%%%%%%%%%%%%%%%%%%%%%%%%%%%%%%%%%%%

\section{Introduction}
\label{sec:intro}

The logic of \emph{quantified Boolean formulas
(QBFs)}~\cite{DBLP:series/faia/BuningB09} extends propositional logic by
existential and universal quantification of propositional
variables. Consequently, the QBF satisfiability problem is
PSPACE-complete~\cite{DBLP:conf/stoc/StockmeyerM73}.  QBF satisfiability is a
restricted form of a \emph{quantified constraint satisfaction problem (QCSP)},
cf.~\cite{DBLP:conf/aaai/BordeauxCM05,DBLP:journals/csur/Chen09,DBLP:conf/birthday/Chen12,DBLP:conf/dagstuhl/Martin17},
where all variables are defined over a Boolean domain.

The \emph{polynomial hierarchy
  (PH)}~\cite{DBLP:conf/focs/MeyerS72,DBLP:journals/tcs/Stockmeyer76,DBLP:journals/tcs/Wrathall76} allows 
to describe the complexity of problems that are beyond the classes P and NP. The satisfiability
problem of a QBF $\psi$ in \emph{prenex
conjunctive normal form (PCNF)} with $k \geq 0$ quantifier alternations is located at
level $k+1$ of PH~\cite{DBLP:journals/tcs/Stockmeyer76,DBLP:journals/tcs/Wrathall76} and either $\Sigma_{k+1}^{P}$-complete or
$\Pi_{k+1}^{P}$-complete, depending on the first quantifier in 
$\psi$. Due to this property,  
practically relevant problems
from any level of PH 
up to the class PSPACE (here 
with arbitrarily nested quantifiers) 
can succinctly be encoded as QBFs.

Efficient solvers are highly
requested to solve QBF encodings of problems.
Competitions like \emph{QBFEVAL} or the \emph{QBF Galleries} have been
driving solver development~\cite{qbflibandeval,gallery14,Lonsing201692}. 
State-of-the-art solvers are based on solving paradigms
like, e.g., 
expansion~\cite{DBLP:conf/fmcad/AyariB02,DBLP:conf/sat/Biere04a,Janota20161}
or Q-resolution~\cite{DBLP:journals/iandc/BuningKF95}. These two
paradigms are \emph{orthogonal} by proof
complexity~\cite{beyersdorff_et_al:LIPIcs:2015:4905,DBLP:journals/tcs/JanotaM15,DBLP:conf/cav/Tentrup17}. Informally,
orthogonal paradigms have complementary strengths on certain families
of formulas.

Motivated by the variety of available QBF solving paradigms and solvers, we
present an \emph{experimental study of the effects of quantifier alternations}
on the evaluation of QBF solvers. To this end, we consider benchmarks,
solvers, and preprocessors from QBFEVAL'17~\cite{qbfeval2017}.  As our main
result, we show that the performance of solvers based on different and,
notably, orthogonal solving paradigms substantially varies depending on the
numbers of alternations.  Instances with a particular number of alternations
may be overrepresented (i.e., appear more frequently) in a benchmark set, thus
resulting in \emph{alternation bias}. In this case, overall solver rankings by
total solved instances may not provide a comprehensive picture as they might
only reflect the strengths of certain solvers on overrepresented instances,
but not the (perhaps orthogonal) strengths of other solvers on
underrepresented ones.

In related work~\cite{DBLP:journals/fuin/MarinNPTG16}, the correlation between
solver performance and various syntactic features such as
treewidth~\cite{DBLP:conf/stacs/AtseriasO13,DBLP:conf/lpar/PulinaT08} was
analyzed. In contrast to that, we do not study such correlations.  By our
study we \emph{a posteriori} highlight diversity of solver performance based
on the single feature of alternations, which are naturally related to the
theoretical hardness of instances in PH.  Recently, alternations have become
of interest also in theoretical work on QBF proof
complexity~\cite{DBLP:conf/innovations/BeyersdorffBH18,DBLP:conf/fsttcs/BeyersdorffHP17,DBLP:journals/toct/Chen17}.

We aim at raising the awareness and importance of quantifier
alternations in comparative studies of QBF solver performance and the
potential negative impact on the progress of QBF solver development.  If
solvers are evaluated on benchmark sets with alternation bias and
alternations are neglected in the analysis, then future research may
inadvertently be narrowed down to only exploring approaches that
perform well on overrepresented instances with a certain number of
alternations. The risk of such detrimental effects on  
a research field driven by empirical analysis has been pointed
already in the early days of propositional satisfiability (SAT)
solving~\cite{DBLP:journals/heuristics/Hooker95} and also with respect
to more recent SAT solver 
competitions~\cite{DBLP:journals/ai/BalintBJS15,DBLP:journals/ai/BalyoBIS16,DBLP:conf/aaai/BalyoHJ17}. In
contrast to the NP-completeness of SAT, the complexity landscape of QBF
encodings defined by PH is more diverse, which gives rise to
several sources of inadvertent convergence of research lines.

In addition to focusing on alternations, we report on \emph{virtual best solver (VBS)}
statistics, where the VBS solved between 50\% and 70\% more instances
than the single overall best solver on a benchmark set. These results
indicate the potential of combining orthogonal QBF proof systems in
solvers. Moreover, we point out that overall low-ranked solvers
potentially solve more instances uniquely and have larger 
contributions to the VBS than high-ranked ones. Similar observations
were made in the context of SAT solver
competitions~\cite{DBLP:conf/sat/XuHHL12}.

The majority of benchmarks in QBFLIB~\cite{qbflibandeval},
the QBF research community portal, has no more than two quantifier
alternations. Hence problems from the first three 
levels in PH have been, and are, of primary interest to practitioners.
However, to strengthen QBF solving as a key technology for solving
problems from \emph{any} levels of PH up to PSPACE-complete problems,
QBF solvers must be improved on instances with
\emph{any} number of alternations. 
Our empirical study 
motivates the development of methods to evaluate orthogonal solving paradigms
by taking quantifier alternations into account. 
This is necessary to showcase the broad range
of existing paradigms for practical QBF applications.

%%%%%%%%%%%%%%%%%%%%%%%%%%%%%%%%%%%%%%%%%%%%%%%%%%%%%%%%%%%%%%%%%%%%%%%%%%%%%%%%
%%%%%%%%%%%%%%%%%%%%%%%%%%%%%%%%%%%%%%%%%%%%%%%%%%%%%%%%%%%%%%%%%%%%%%%%%%%%%%%%

\section{Preliminaries}
\label{sect:setup:prelims}

We consider QBFs $\psi := \Pi.\phi$ in \emph{prenex conjunctive normal
  form (PCNF)} consisting of a \emph{quantifier prefix} $\Pi := Q_1B_1
\ldots Q_nB_n$ and a quantifier-free propositional formula $\phi$ in
\emph{CNF}. A CNF consists of a conjunction of \emph{clauses}. A clause is a
disjunction of \emph{literals}. A literal is either a propositional variable
$x$ or its negation $\neg x$. The prefix $\Pi$ is a linearly ordered sequence of \emph{quantifier
  blocks (qblocks)} $Q_iB_i$, where $Q_i \in \{\forall,\exists\}$ is a
quantifier and $B_i$ is a block (i.e., a set) of propositional
variables with $B_i \cap B_j = \emptyset$ for $i \not = j$. 
The notation $Q_iB_i$ is shorthand for $Q_ix_1 \ldots
Q_ix_m$ for all $x_j \in B_i$. Formula $\phi$ is defined precisely
over the variables that appear in $\Pi$. 
If $Q_i
= Q_{i+1}$ then $B_i$ and $B_{i+1}$ are merged to obtain $Q_i(B_{i}
\cup B_{i+1})$. Hence adjacent qblocks are
quantified differently. 
Without loss of generality, we assume that the innermost quantifier $Q_n = \exists$ is 
existential. (If $Q_n = \forall$ then $B_n$
is eliminated  by \emph{universal
  reduction}~\cite{DBLP:journals/iandc/BuningKF95}.)  A PCNF with $n$
qblocks has $n-1$ \emph{quantifier alternations}.

The \emph{semantics} of PCNFs are defined recursively. 
The PCNF consisting only of the \emph{syntactic truth constant} $\top$
($\bot$) is satisfiable (unsatisfiable).  A PCNF $\psi := Q_1B_1 \ldots
Q_nB_n.\phi$ with $Q_1 = \exists$ ($Q_1 = \forall$) is satisfiable iff, for $x
\in B_1$, $\psi[x]$ or (and) $\psi[\neg x]$ is satisfiable, where $\psi[x]$
($\psi[\neg x]$) is the PCNF obtained from $\psi$ by replacing all occurrences
of $x$ by $\top$ ($\bot$) and deleting $x$ from $B_1$.

To make the presentation of our experimental study self-contained, we
introduce \emph{QBF proof systems} only informally and refer to
a standard, formal definition of propositional proof systems~\cite{DBLP:journals/jsyml/CookR79}.  A \emph{QBF proof system}
$\mathcal{PS}$ is a formal system consisting of \emph{inference rules}. The
inference rules allow to derive new formulas (e.g.~clauses) from a given QBF
$\psi$ and from previously derived formulas.  A QBF proof system
$\mathcal{PS}$ is \emph{correct} if, for any QBF $\psi$, it holds that if the
formula $\bot$ (\emph{false}, e.g., the empty clause) is derivable in
$\mathcal{PS}$ from $\psi$ then $\psi$ is
unsatisfiable.\footnote{Theoretical work on QBF proof systems typically
  focuses on unsatisfiable QBFs.}  A QBF proof system
$\mathcal{PS}$ is \emph{complete} if, for any QBF $\psi$, it holds that if
$\psi$ is unsatisfiable then $\bot$ is derivable in $\mathcal{PS}$ from
$\psi$. 
A \emph{proof} $P$ of an unsatisfiable QBF $\psi$ in $\mathcal{PS}$ is a
sequence of given formulas and formulas derived by inference rules ending in
$\bot$. The \emph{length} $|P|$ of a proof $P$ is the sum of the sizes of all
formulas in $P$.

Let $\mathcal{PS}$ and $\mathcal{PS'}$ be QBF proof systems and $\Psi$ be a
family of unsatisfiable QBFs. Let $P$ be a proof of some QBF $\psi \in \Psi$ in
$\mathcal{PS}$ such that the length $|P|$ of $P$ is polynomial in the size of
$\psi$. Assume that the length $|P'|$ of every proof $P'$ of $\psi \in \Psi$
in $\mathcal{PS'}$ is exponential in the size of $\psi$. Then $\mathcal{PS}$
is \emph{stronger} than $\mathcal{PS}'$ with respect to family
$\Psi$. Two QBF proof systems $\mathcal{PS}$ and $\mathcal{PS'}$ are \emph{orthogonal} if
$\mathcal{PS}$ is stronger than $\mathcal{PS}'$ with respect to a family
$\Psi$ and $\mathcal{PS}'$ is stronger than $\mathcal{PS}$ with respect to
some other family $\Psi'$. The relation between QBF proof systems in terms of
their strengths is studied in the research field of \emph{QBF proof complexity}. 

QBF proof systems are the formal foundation of QBF solver implementations. Expansion~\cite{DBLP:conf/fmcad/AyariB02,DBLP:conf/sat/Biere04a,Janota20161}
and Q-resolution~\cite{DBLP:journals/iandc/BuningKF95} are traditional QBF
proof systems that are
orthogonal~\cite{beyersdorff_et_al:LIPIcs:2015:4905,DBLP:journals/tcs/JanotaM15,DBLP:conf/cav/Tentrup17}. Orthogonal
proof systems are of particular interest for practical QBF solving since they give rise
to solvers that have individual, complementary strengths on certain families
of formulas. In our experiments, we highlight the potential of
combining orthogonal proof systems in QBF solvers.

%%%%%%%%%%%%%%%%%%%%%%%%%%%%%%%%%%%%%%%%%%%%%%%%%%%%%%%%%%%%%%%%%%%%%%%%%%%%%%%%
%%%%%%%%%%%%%%%%%%%%%%%%%%%%%%%%%%%%%%%%%%%%%%%%%%%%%%%%%%%%%%%%%%%%%%%%%%%%%%%%

\section{Experimental Setup}
\label{sect:setup:experiments}

For our experimental study we use the set $\instancesnopreallNEW$ containing 
523 PCNFs from 
\mbox{QBFEVAL'17}~\cite{qbfeval2017}. Partitioning $\instancesnopreallNEW$ by
numbers of qblocks results in 64
classes. Table~\ref{fig:exp:825:noprepro:histogram} shows a histogram
of $\instancesnopreallNEW$ by the numbers of formulas (\#f) in 
\begin{wraptable}{r}{0.24\textwidth}
\vspace{-0.75cm}
\begin{tabular}{l@{\hspace*{0.2cm}}r@{\hspace*{0.2cm}}r}
\hline
\#q & \#f & $\#\textnormal{f}_{\textnormal{L}}$\\
\hline
1      &   0  & 253   \\
2      &  90   & 7,319  \\  
3      &  236   & 4,110  \\  
4--10    &  70  &  2,185 \\ 
11--20    &  42  &  437 \\ 
21--   &   85  &  2,444 \\
\hline
1--3    & 326  & 11,682 \\
4--     & 197  &  5,066 \\
\hline
\end{tabular}
\vspace{-0.4cm}
\caption{}
\label{fig:exp:825:noprepro:histogram}
\vspace{-0.5cm}
\end{wraptable}
classes
defined by the number of qblocks (\#q).  Instances with up to three
qblocks (row ``1--3'') amount to 62\% of all instances and hence are
overrepresented in $\instancesnopreallNEW$. 
To generate
$\instancesnopreallNEW$, instances were sampled from instance
categories in QBFLIB in addition to newly submitted ones based on
empirical hardness results from previous competitions.  We also computed a
histogram of a QBFLIB snapshot containing 16,748 instances (column
$\#\textnormal{f}_{\textnormal{L}}$ in
Table~\ref{fig:exp:825:noprepro:histogram}). Instances with no more 
than three qblocks (row ``1--3'') are also overrepresented (69\%) in
that snapshot. Hence alternation bias in $\instancesnopreallNEW$ follows from a related bias in QBFLIB, 
which is due to the focus of QBF practitioners on problems located at low
levels in PH. Moreover, the bias does \emph{not} result from a flawed
selection of competition instances. We use the terminology
``overrepresented'' and ``bias'' for the statistical fact that instances with few
qblocks appear more frequently in $\instancesnopreallNEW$.

In order to evaluate the impact of qblocks on solver performance, we
consider~11  solvers that participated in
QBFEVAL'17 and were
top-ranked.\footnote{For some solvers where version numbers are not reported,
  the authors kindly provided us with the
  competition versions, which were not publicly
  available. We excluded the solver \aigsolve because we observed
  assertion failures on certain instances.}  
The solvers implement the following six different \emph{solving
  paradigms:}

\begin{enumerate}
\item
  \emph{Expansion}~\cite{DBLP:conf/fmcad/AyariB02,DBLP:conf/sat/Biere04a}
  eliminates variables from a PCNF $\psi$ until the formula reduces to
  either \emph{true} or \emph{false}.  \rareqs~1.1~\cite{Janota20161}
  applies recursive expansion based on \emph{counterexample-guided
    abstraction refinement
    (CEGAR)}~\cite{DBLP:journals/jacm/ClarkeGJLV03}, while
  \ijtihad operates in a
  non-recursive way.
  \revqfun~0.1~\cite{DBLP:journals/corr/JanotaQFUN2017} extends \rareqs by
  machine learning techniques, and
  \dynqbf~\cite{DBLP:conf/aiia/CharwatW17} exploits QBF tree
    decompositions. Theoretical properties of expansion as a proof
    system, which underlies implementations of expansion solvers, have been \nolinebreak intensively
    \nolinebreak studied~\cite{beyersdorff_et_al:LIPIcs:2015:4905,DBLP:journals/tcs/JanotaM15}.
  
\label{paradigm:exp}

\item \emph{QDPLL}~\cite{DBLP:conf/aaai/CadoliGS98} is a backtracking
  search procedure that generalizes the DPLL
  algorithm~\cite{DBLP:journals/cacm/DavisLL62}.
  \ghostqcegar~\cite{Janota20161,DBLP:conf/sat/KlieberSGC10} combines
  QDPLL with clause and cube learning (a cube is a conjunction of
  literals) based on the \emph{Q-resolution proof
    system}~\cite{DBLP:journals/iandc/BuningKF95}. Additionally, it
  reconstructs the structure of PCNFs encoded by Tseitin
  translation~\cite{Tseitin}, and applies CEGAR-based learning.

\label{paradigm:ghostq}

\item \emph{Nested SAT solving} uses one SAT solver per qblock in
  a PCNF, where universal quantification is handled as negated
  existential quantification. The solver
  \qstsdefnobreaksym~\cite{DBLP:conf/sat/0001JT16,BeyonNPNestedSATSolvers}
  combines nested SAT solving with  structure
  reconstruction. Propositional resolution is the proof system that
  underlies SAT solving.

\label{paradigm:qsts}

\item \emph{Clause selection} and \emph{clausal abstraction} as implemented in
  the solvers \qesto~1.0~\cite{DBLP:conf/ijcai/JanotaM15} and
  \caqe~\cite{DBLP:conf/fmcad/RabeT15,DBLP:conf/cav/Tentrup17},
  respectively, decompose the given PCNF into a sequence of
  propositional formulas and apply CEGAR techniques. The proof system
  implemented in \caqe has been presented
  recently~\cite{DBLP:conf/cav/Tentrup17}.

\label{paradigm:caqe}

\item \emph{Backtracking search with clause and cube learning
  (QCDCL)}~\cite{DBLP:conf/aaai/GiunchigliaNT02,DBLP:journals/jair/GiunchigliaNT06,DBLP:conf/tableaux/Letz02,DBLP:conf/cp/ZhangM02}
  based on Q-resolution 
  extends the CDCL approach for SAT
  solving~\cite{DBLP:series/faia/SilvaLM09} to QBFs. The solver
  \depqbfprefixopt~\cite{DBLP:conf/cade/LonsingE17} implements QCDCL with
  generalized Q-resolution axioms allowing for a stronger calculus to
  derive learned clauses and cubes. \quterandom~\cite{DBLP:conf/sat/PeitlSS17} learns variable dependencies lazily in a run.

\label{paradigm:depqbf}

\item \heretic is based on a \emph{hybrid approach} that combines
  expansion and QCDCL in a sequential portfolio style. Thereby, the
  QCDCL solver \depqbfprefixopt is applied to learn clauses from the
  given QBF, which are then heuristically added to the expansion
  solver \ijtihad.

\label{paradigm:heretic}

\end{enumerate}

%%%%%%%%%%%%%%%%%%%%%%%%%%%%%%%%%%%%%%%%%%%%%%%%%%%%%%%%%%%%%%%%%%%%%%%%%%%%%%%%
%%%%%%%%%%%%%%%%%%%%%%%%%%%%%%%%%%%%%%%%%%%%%%%%%%%%%%%%%%%%%%%%%%%%%%%%%%%%%%%%

\section{Experimental Results}
\label{sect:experiments}

We illustrate a substantial performance diversity of
the above solvers from QBFEVAL'17 on instances with different numbers
of quantifier alternations.  To this end, we rank solvers based on
\emph{instance classes} given by numbers of qblocks similar to
Table~\ref{fig:exp:825:noprepro:histogram}.  Our empirical results are
consistent on instances with and without preprocessing by the state-of-the-art 
tools \bloqqer~\cite{DBLP:journals/jair/HeuleJLSB15} and
\hqspre~\cite{DBLP:conf/tacas/WimmerRM017}. 
Alternation bias
in original instances is present also in preprocessed ones.
Unless stated otherwise, all experiments were run on 
Intel Xeon CPUs (E5-2650v4, 2.20 GHz) with Ubuntu 16.04.1 using CPU
time and memory limits of 1800 seconds and seven GB. Exceeding the
memory limit is counted as a time out.

It is well known that preprocessing may have positive
effects on the performance of certain solvers while negative effects
on others (cf.~\cite{Lonsing201692,DBLP:journals/fuin/MarinNPTG16}).
To compensate for these effects, we applied preprocessing both to
filter the original benchmark set $\instancesnopreallNEW$ and to
preprocess instances. Many preprocessing techniques used to simplify a
QBF by eliminating clauses and literals 
are restricted variants of solving approaches, hence instances might be solved
already by preprocessing.

We ran \bloqqer (version 37) with a time limit of two
hours as a filter on set $\instancesnopreallNEW$ to obtain the set
$\instancesnoprefilteredNEW$ containing 437 \emph{original} PCNFs, where we discarded
76 instances from $\instancesnopreallNEW$ that were solved already by
\bloqqer and ten instances that became propositional, i.e., which
ended up having a single quantifier block of existential variables
only. 
\bloqqer exceeded the time limit on 39 instances, which we
included in their original form in set $\instancesnoprefilteredNEW$. 

In a similar way, we filtered set $\instancesnopreallNEW$ using
\hqspre to obtain the set $\instancesnoprefilteredHQSpreNEW$
containing 312 \emph{original} instances, where we discarded 183 instances solved by
\hqspre and 28 which became propositional, and we included 42 original
ones in $\instancesnoprefilteredHQSpreNEW$ where \hqspre exceeded the
resource limits. 
We did not consider a variant of \hqspre that applies a restricted
form of preprocessing to preserve gate structure present in
formulas~\cite{DBLP:conf/tacas/WimmerRM017}. Compared to the unrestricted 
variant of \hqspre we used, the restricted one did not
improve overall solver performance. 

By applying \bloqqer and \hqspre to the filtered sets $\instancesnoprefilteredNEW$ and
$\instancesnoprefilteredHQSpreNEW$ 
again, we generated the sets $\instancesprefilteredNEW$ and
$\instancesprefilteredHQSpreNEW$, respectively, containing
preprocessed instances and those original instances where the
preprocessors exceeded the resource limits.  We disabled any
additional use of \bloqqer or \hqspre as separate preprocessing
modules integrated in some solvers. \emph{In the following, we focus
  our analysis on the four sets $\instancesnoprefilteredNEW$,
  $\instancesprefilteredNEW$, $\instancesnoprefilteredHQSpreNEW$, and
  $\instancesprefilteredHQSpreNEW$.}

\begin{figure}[t]
\begin{minipage}[b]{0.48\textwidth}
\begin{center}
\subfloat[Sets $\instancesnoprefilteredNEW$ (x-axis) and $\instancesprefilteredNEW$.]{
\includegraphics[scale=0.75]{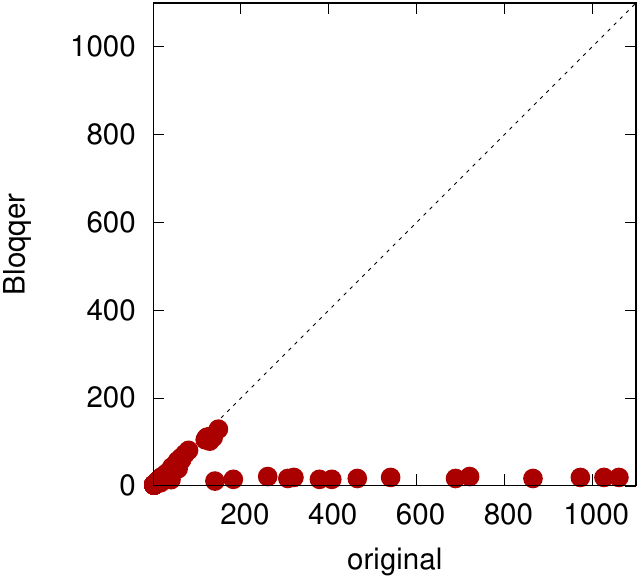}
\label{fig:exp:scatter:bloqqer}
}%end subfloat
\end{center}
\end{minipage}
\hfill
\begin{minipage}[b]{0.48\textwidth}
\begin{center}
\subfloat[Sets $\instancesnoprefilteredHQSpreNEW$ (x-axis) and $\instancesprefilteredHQSpreNEW$.]{
\includegraphics[scale=0.75]{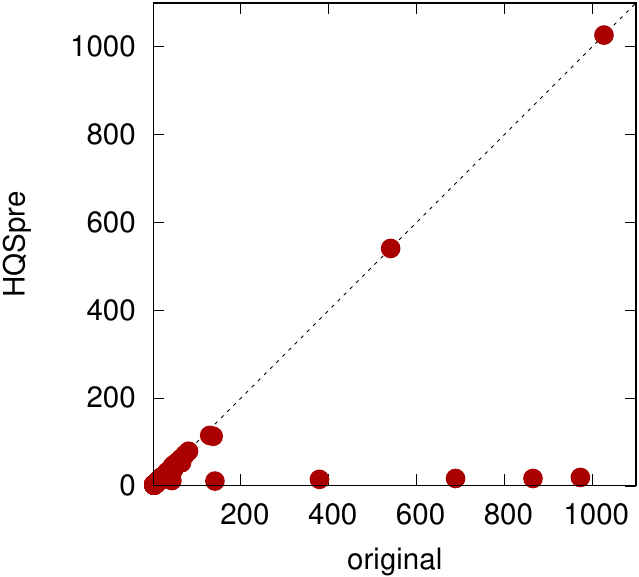}
\label{fig:exp:scatter:hqspre}
}%end subfloat
\end{center}
\end{minipage}
\caption{Numbers of qblocks before (``original'') and after preprocessing by
  \bloqqer (\ref{fig:exp:scatter:bloqqer}) and \hqspre
  (\ref{fig:exp:scatter:hqspre}) on filtered (x-axes) and preprocessed instances (y-axes), respectively.}
\label{fig:exp:scatter}
\end{figure}

%%%%%%%%%%%%%%%%%%%%%%%%%%%%%%%%%
\begin{table}[t]
\caption{Histograms of the benchmark sets
  $\instancesnoprefilteredNEW$ and 
  $\instancesprefilteredNEW$ (filtered and preprocessed by \bloqqer), and
  $\instancesnoprefilteredHQSpreNEW$ and
  $\instancesprefilteredHQSpreNEW$ (filtered and preprocessed by \hqspre) illustrating the numbers of formulas (\#f) in
classes given by the number of qblocks (\#q).}
\label{fig:exp:benchmark:sets:histograms}
\addtocounter{table}{-1}
\begin{minipage}[b]{0.24\textwidth}
\begin{center}
\subfloat[Set~$\instancesnoprefilteredNEW$.]{
\begin{tabular}{l@{\hspace*{0.5cm}}r}
\hline
\#q & \#f \\
\hline
2 & 63   \\
3 & 215   \\
4--10 & 63   \\
11--20 & 36  \\
21-- & 60  \\
\hline
2--3 & 278   \\
4-- & 159  \\
\hline
\end{tabular}
\label{fig:exp:benchmark:sets:histograms:0}
}% end: subfloat
\end{center}
\end{minipage}
\begin{minipage}[b]{0.24\textwidth}
\begin{center}
\subfloat[Set $\instancesprefilteredNEW$.]{
\begin{tabular}{l@{\hspace*{0.5cm}}r}
\hline
\#q & \#f \\
\hline
2 & 65  \\
3 & 218  \\
4--10 & 59  \\
11-20 & 53  \\
21-- & 42  \\
\hline
2--3 & 283  \\
4-- & 154  \\
\hline
\end{tabular}
\label{fig:exp:benchmark:sets:histograms:1}
}% end: subfloat
\end{center}
\end{minipage}
\begin{minipage}[b]{0.24\textwidth}
\begin{center}
\subfloat[Set $\instancesnoprefilteredHQSpreNEW$.]{
\begin{tabular}{l@{\hspace*{0.5cm}}r}
\hline
\#q & \#f \\
\hline
2 & 70   \\
3 & 145   \\
4--10 & 26  \\
11--20 & 30   \\
21-- & 41   \\
\hline
2--3 & 215   \\
4-- & 97 \\
\hline
\end{tabular}
\label{fig:exp:benchmark:sets:histograms:2}
}% end: subfloat
\end{center}
\end{minipage}
\begin{minipage}[b]{0.24\textwidth}
\begin{center}
\subfloat[Set $\instancesprefilteredHQSpreNEW$.]{
\begin{tabular}{l@{\hspace*{0.5cm}}r}
\hline
\#q & \#f \\
\hline
2 & 70  \\
3 & 145  \\
4--10 & 26  \\
11--20 & 40  \\
21-- & 31  \\
\hline
2--3 & 215  \\
4-- & 97  \\
\hline
\end{tabular}
\label{fig:exp:benchmark:sets:histograms:3}
}% end: subfloat
\end{center}
\end{minipage}
\refstepcounter{table}
\end{table}

%%%%%%%%%%%%%%%%%%%%%%%%%%%%%%%%%

The application of \bloqqer and \hqspre to sets
$\instancesnoprefilteredNEW$ and $\instancesnoprefilteredHQSpreNEW$
reduces the number of qblocks in instances considerably. This is 
illustrated by the scatter plots in
Figures~\ref{fig:exp:scatter:bloqqer}
and~\ref{fig:exp:scatter:hqspre}, respectively.  The average number of
qblocks decreases from 29 in set $\instancesnoprefilteredNEW$ to 10 in
set $\instancesprefilteredNEW$. Likewise, the average decreases from~24 
in set $\instancesnoprefilteredHQSpreNEW$ to 14 in set
$\instancesprefilteredHQSpreNEW$. As an extreme case, the number of
qblocks in an instance in $\instancesnoprefilteredNEW$ was reduced by
\bloqqer from 1061 to 19.

In all sets
$\instancesnoprefilteredNEW$, $\instancesprefilteredNEW$,
$\instancesnoprefilteredHQSpreNEW$, and
$\instancesprefilteredHQSpreNEW$, the median number of qblocks is
three. This is due to 
alternation bias like in the original set
$\instancesnopreallNEW$
(Table~\ref{fig:exp:825:noprepro:histogram}). The related 
histograms are shown in
Tables~\ref{fig:exp:benchmark:sets:histograms:0}
to~\ref{fig:exp:benchmark:sets:histograms:3},
where instances with no more than three qblocks are overrepresented
(rows ``2--3'') as they amount to between 63\% and 68\% of all 437, respectively, 312 
instances. 
Set $\instancesnoprefilteredNEW$ has 59 classes by numbers
of qblocks compared to 45 in set $\instancesprefilteredNEW$, and set
$\instancesnoprefilteredHQSpreNEW$ has 42 compared to 40 in set
$\instancesprefilteredHQSpreNEW$.  \bloqqer reduces the number of
instances with 21 or more qblocks (lines ``21--'') from 60 in
$\instancesnoprefilteredNEW$ to 42 in
$\instancesprefilteredNEW$ (Tables~\ref{fig:exp:benchmark:sets:histograms:0}
and~\ref{fig:exp:benchmark:sets:histograms:1}). \hqspre reduces this number from
41 in $\instancesnoprefilteredHQSpreNEW$ to 31 in
$\instancesprefilteredHQSpreNEW$ (Tables~\ref{fig:exp:benchmark:sets:histograms:2}
\nolinebreak and~\ref{fig:exp:benchmark:sets:histograms:3}).

\begin{table}[t]
\caption{Solvers and corresponding paradigms (\emph{P}) from Section~\ref{sect:setup:experiments}, solved instances (\emph{S}), unsatisfiable
(\emph{$\bot$}) and satisfiable ones (\emph{$\top$}), total CPU 
time including time outs, and uniquely solved instances (\emph{U}) on sets $\instancesnoprefilteredNEW$
(\ref{fig:exp:all:sets:filtered:bloqqer}), \mbox{$\instancesprefilteredNEW$ (\ref{fig:exp:all:sets:prepro:bloqqer}), $\instancesnoprefilteredHQSpreNEW$
(\ref{fig:exp:all:sets:filtered:hqspre}), and $\instancesprefilteredHQSpreNEW$ (\ref{fig:exp:all:sets:prepro:hqspre}).}}
\addtocounter{table}{-1}
\begin{minipage}[b]{0.50\textwidth}
\begin{center}
\subfloat[Set $\instancesnoprefilteredNEW$ filtered by \bloqqer.]{
{\setlength\tabcolsep{0.125cm}
\begin{tabular}{lrrrrrr}
\hline
\emph{Solver} & \multicolumn{1}{c}{\emph{P}} & \multicolumn{1}{c}{\emph{S}} &
\multicolumn{1}{c}{\emph{$\bot$}} & \multicolumn{1}{c}{\emph{$\top$}} &
\emph{Time} & \emph{U}\\
\hline
\revqfun & \ref*{paradigm:exp} & 174 & 106 & 68 & 497K & 6 \\ 
\ghostqcegar & \ref*{paradigm:ghostq} & 145 & 79 & 66 & 547K & 12 \\ 
\rareqs & \ref*{paradigm:exp} & 126 & 94 & 32 & 577K & 4 \\ 
\caqe & \ref*{paradigm:caqe} & 126 & 87 & 39 & 578K & 6 \\ 
\heretic & \ref*{paradigm:heretic} & 122 & 95 & 27 & 580K & 0\\ 
\depqbfprefixopt & \ref*{paradigm:depqbf} & 115 & 78 & 37 & 603K & 16 \\ 
\ijtihad & \ref*{paradigm:exp} & 110 & 88 & 22 & 599K & 1 \\ 
\qstsdefnobreaksym & \ref*{paradigm:qsts} & 103 & 75 & 28 & 618K & 3 \\ 
\quterandom & \ref*{paradigm:depqbf} & 77 & 47 & 30 & 658K & 0 \\ 
\qesto & \ref*{paradigm:caqe} & 76 & 56 & 20 & 661K & 0 \\ 
\dynqbf & \ref*{paradigm:exp} & 47 & 27 & 20 & 714K & 9 \\ 
\hline
\end{tabular}
}%end: setlength
\label{fig:exp:all:sets:filtered:bloqqer}
}%end subfloat
\end{center}
\end{minipage}
\hfill
\begin{minipage}[b]{0.50\textwidth}
\begin{center}
\subfloat[\mbox{Set $\instancesprefilteredNEW$ preprocessed by \bloqqer.}]{
{\setlength\tabcolsep{0.125cm}
\begin{tabular}{lrrrrrr}
\hline
\emph{Solver} & \multicolumn{1}{c}{\emph{P}} & \multicolumn{1}{c}{\emph{S}} &
\multicolumn{1}{c}{\emph{$\bot$}} & \multicolumn{1}{c}{\emph{$\top$}} &
\emph{Time} & \emph{U}\\
\hline
\rareqs & \ref*{paradigm:exp} & 175 & 127 & 48 & 499K & 5 \\ 
\caqe & \ref*{paradigm:caqe} & 169 & 114 & 55 & 514K  & 0 \\ 
\heretic & \ref*{paradigm:heretic} & 164 & 119 & 45 & 513K  & 0\\ 
\ijtihad & \ref*{paradigm:exp} & 136 & 103 & 33 & 555K  & 2 \\ 
\revqfun & \ref*{paradigm:exp} & 135 & 92 & 43 & 563K  & 3 \\ 
\qstsdefnobreaksym & \ref*{paradigm:qsts} & 127 & 98 & 29 & 576K  & 12\\ 
\qesto & \ref*{paradigm:caqe} & 115 & 84 & 31 & 601K  & 1 \\ 
\depqbfprefixopt & \ref*{paradigm:depqbf} & 102 & 64 & 38 & 624K  & 3\\ 
\ghostqcegar & \ref*{paradigm:ghostq} & 82 & 47 & 35 & 661K  & 1\\ 
\quterandom & \ref*{paradigm:depqbf} & 73 & 56 & 17 & 672K  & 0 \\ 
\dynqbf & \ref*{paradigm:exp} & 65 & 37 & 28 & 684K  & 25 \\ 
\hline
\end{tabular}
}%end: setlength
\label{fig:exp:all:sets:prepro:bloqqer}
}%end subfloat
\end{center}
\end{minipage}

\medskip

\begin{minipage}[b]{0.50\textwidth}
\begin{center}
\subfloat[Set $\instancesnoprefilteredHQSpreNEW$ filtered by \hqspre.]{
{\setlength\tabcolsep{0.125cm}
\begin{tabular}{lrrrrrr}
\hline
\emph{Solver} & \multicolumn{1}{c}{\emph{P}} & \multicolumn{1}{c}{\emph{S}} &
\multicolumn{1}{c}{\emph{$\bot$}} & \multicolumn{1}{c}{\emph{$\top$}} &
\emph{Time} & \emph{U} \\
\hline
\ghostqcegar & \ref*{paradigm:ghostq} & 112 & 61 & 51 & 373K & 15 \\ 
\revqfun & \ref*{paradigm:exp} & 110 & 58 & 52 & 376K  & 6 \\ 
\caqe & \ref*{paradigm:caqe} & 68 & 42 & 26 & 454K  & 6 \\ 
\depqbfprefixopt & \ref*{paradigm:depqbf} & 64 & 41 & 23 & 461K & 4 \\ 
\qstsdefnobreaksym & \ref*{paradigm:qsts} & 56 & 34 & 22 & 470K & 3 \\ 
\rareqs & \ref*{paradigm:exp} & 50 & 34 & 16 & 482K & 1 \\ 
\heretic & \ref*{paradigm:heretic} & 49 & 34 & 15 & 485K & 0 \\ 
\quterandom & \ref*{paradigm:depqbf} & 47 & 25 & 22 & 486K & 0 \\ 
\dynqbf & \ref*{paradigm:exp} & 46 & 24 & 22 & 488K & 9 \\ 
\qesto & \ref*{paradigm:caqe} &  45 & 30 & 15 & 491K & 0 \\ 
\ijtihad & \ref*{paradigm:exp} & 36 & 27 & 9 & 504K & 1 \\
\hline
\end{tabular}
}%end: setlength
\label{fig:exp:all:sets:filtered:hqspre}
}%end subfloat
\end{center}
\end{minipage}
\hfill
\begin{minipage}[b]{0.50\textwidth}
\begin{center}
\subfloat[\mbox{Set $\instancesprefilteredHQSpreNEW$ preprocessed by \hqspre.}]{
{\setlength\tabcolsep{0.125cm}
\begin{tabular}{lrrrrrr}
\hline
\emph{Solver} & \multicolumn{1}{c}{\emph{P}} & \multicolumn{1}{c}{\emph{S}} & \multicolumn{1}{c}{\emph{$\bot$}} & \multicolumn{1}{c}{\emph{$\top$}} & \emph{Time} & \emph{U} \\
\hline
\caqe & \ref*{paradigm:caqe} & 114 & 65 & 49 & 378K & 6 \\ 
\rareqs & \ref*{paradigm:exp} & 103 & 63 & 40 & 390K & 3 \\ 
\qesto & \ref*{paradigm:caqe} & 97 & 63 & 34 & 402K & 1 \\ 
\revqfun & \ref*{paradigm:exp} & 90 & 57 & 33 & 414K & 6 \\ 
\heretic & \ref*{paradigm:heretic} & 87 & 55 & 32 & 424K & 0 \\ 
\qstsdefnobreaksym & \ref*{paradigm:qsts} & 72 & 46 & 26 & 448K & 1\\ 
\depqbfprefixopt & \ref*{paradigm:depqbf} & 72 & 44 & 28 & 451K & 5\\ 
\quterandom & \ref*{paradigm:depqbf} & 70 & 42 & 28 & 449K & 2 \\ 
\ijtihad & \ref*{paradigm:exp} & 58 & 43 & 15 & 465K & 1 \\ 
\ghostqcegar & \ref*{paradigm:ghostq} & 58 & 33 & 25 & 475K & 0 \\ 
\dynqbf & \ref*{paradigm:exp} & 45 & 24 & 21 & 487K & 17 \\ 
\hline
\end{tabular}
}%end: setlength
\label{fig:exp:all:sets:prepro:hqspre}
}%end subfloat
\end{center}
\end{minipage}
\label{fig:exp:all:sets}
\refstepcounter{table}
\end{table}

\begin{table}[t]
\caption{Solvers and corresponding solving paradigms (\emph{P}) as listed in
Section~\ref{sect:setup:experiments}, solved instances (\emph{S},
cf.~Tables~\ref{fig:exp:all:sets:filtered:bloqqer}
to~\ref{fig:exp:all:sets:prepro:hqspre}), average ($\overline{q}$) and median
number ($\tilde{q}$) of qblocks of respective solved instances in the
considered benchmark sets. Rows ``$\bigcup$'' show statistics for the total
number of instances solved by any solver based on a particular paradigm.}
\label{fig:exp:all:sets:stats}
\begin{center}
{\setlength\tabcolsep{0.15cm}
\begin{tabular}{llrrr|rrr|rrr|rrr}
\hline
 & & \multicolumn{3}{c|}{$\instancesnoprefilteredNEW$} &
\multicolumn{3}{c|}{$\instancesprefilteredNEW$} &
\multicolumn{3}{c|}{$\instancesnoprefilteredHQSpreNEW$} & \multicolumn{3}{c}{$\instancesprefilteredHQSpreNEW$} \\
\hline
\multicolumn{1}{c}{\emph{P}} & \emph{Solver} & \multicolumn{1}{c}{\emph{S}} & \multicolumn{1}{c}{\emph{$\overline{q}$}} &
\multicolumn{1}{c|}{\emph{$\tilde{q}$}} & \multicolumn{1}{c}{\emph{S}} &
\multicolumn{1}{c}{\emph{$\overline{q}$}} &
\multicolumn{1}{c|}{\emph{$\tilde{q}$}} & \multicolumn{1}{c}{\emph{S}} &
\multicolumn{1}{c}{\emph{$\overline{q}$}} &
\multicolumn{1}{c|}{\emph{$\tilde{q}$}} & \multicolumn{1}{c}{\emph{S}} &
\multicolumn{1}{c}{\emph{$\overline{q}$}} &
\multicolumn{1}{c}{\emph{$\tilde{q}$}} \\
\hline
\multirow{5}{*}{\ref*{paradigm:exp}} & \dynqbf & 47 & 6.1  & 3.0 &  65 & 9.0 & 3.0   & 46 & 4.8 & 3.0 & 45 & 3.3 & 2.0 \\ 
 & \ijtihad & 110 & 42.1 & 5.0 &       136 & 12.7 & 3.0  & 36 & 40.5 & 3.0 &  58 & 17.6 & 3.0 \\ 
 & \rareqs & 126 & 39.8 & 3.0 &         175 & 11.2 & 3.0   & 50& 22.6 & 3.0 &  103 & 11.5 & 3.0 \\ 
 & \revqfun  & 174 & 55.1 & 3.0 &       135 & 12.5 & 3.0  & 110 & 47.4 & 3.0 &  90 & 24.0 & 3.0 \\ 
 & $\bigcup$  & 228 & 45.9 & 3.0 &     238   & 9.6 & 3.0  & 145 & 37.8 & 3.0 & 150  & 16.6 & 3.0 \\ 
\hline
 \ref*{paradigm:ghostq} & \ghostqcegar & 145 & 12.5 & 3.0 & 82 & 15.8 & 3.0  &  112 & 7.5 & 3.0 & 58 & 8.1 & 3.0 \\ 
\hline
\ref*{paradigm:qsts} & \qstsdefnobreaksym  & 103 & 63.2 & 5.0 & 127 & 15.6 & 5.0 & 56 & 65.3 & 3.0 & 72 & 22.6 & 3.0\\ 
\hline
\multirow{3}{*}{\ref*{paradigm:caqe}} & \caqe & 126 & 44.3 & 5.0 &    169 & 12.9 & 3.0 & 68 & 37.4 & 3.0 & 114 & 12.0 & 3.0 \\ 
 & \qesto  & 76 & 47.7 & 3.0 &    115 & 15.5 & 3.0 & 45 & 15.6 & 3.0 & 97 &  8.1 & 3.0 \\ 
 & $\bigcup$  & 134 & 41.9 & 3.5 &   182     & 12.5 & 3.0  & 74 & 34.7 & 3.0 & 127  & 11.6 & 3.0 \\ 
\hline
\multirow{3}{*}{\ref*{paradigm:depqbf}} & \depqbfprefixopt  & 115 & 45.7 & 5.0 & 102 & 17.8 & 8.5 & 64 & 21.2 & 8.0 & 72 & 10.5 & 3.0 \\ 
 & \quterandom  & 77 & 30.0 & 4.0 & 73 & 20.7 & 9.0 & 47 & 16.4 & 3.0 & 70 &  9.7 & 3.0 \\ 
 & $\bigcup$  & 137 & 38.8 & 3.0 &    117    & 16.2 & 6.0  & 83 & 17.0 & 3.0 & 97  & 9.2 & 3.0 \\ 
\hline
\ref*{paradigm:heretic} & \heretic  & 122 & 39.5 & 5.0 & 164 & 12.5 & 5.0 & 49 & 34.4 & 3.0 & 87 & 14.1 & 3.0 \\ 
\hline
\end{tabular}
}%end: setlength
\end{center}
\end{table}

\subsection{Solved Instances: Overall Rankings}

We first analyze overall solver performance by ranking solvers according to
total numbers of instances 
solved in the benchmark sets $\instancesnoprefilteredNEW$,
$\instancesprefilteredNEW$, $\instancesnoprefilteredHQSpreNEW$, and
$\instancesprefilteredHQSpreNEW$. Then we show that the strengths of certain
solvers and solving paradigms are not reflected in such overall rankings. To
highlight these individual strengths, in Section~\ref{sec:class:analysis}
below we carry out a more fine-grained
analysis of solver performance based on
instances that were solved in instance classes defined by their number of
qblocks. Our results show that there is a considerable performance diversity
between solvers and solving paradigms with respect to classes.

Tables~\ref{fig:exp:all:sets:filtered:bloqqer}
to~\ref{fig:exp:all:sets:prepro:hqspre} show overall solver rankings by total
numbers of solved instances. Solver performance greatly varies depending on 
preprocessing. For example, while \rareqs, \caqe, and \qesto clearly benefit
from preprocessing, it is harmful for \ghostqcegar and \revqfun. 
The expansion solvers \rareqs and \revqfun (paradigm~\ref*{paradigm:exp}) 
dominate the rankings on sets
$\instancesnoprefilteredNEW$ and $\instancesprefilteredNEW$
(Tables~\ref{fig:exp:all:sets:filtered:bloqqer}
and~\ref{fig:exp:all:sets:prepro:bloqqer}), and are ranked second on sets
$\instancesnoprefilteredHQSpreNEW$ and $\instancesprefilteredHQSpreNEW$
(Tables~\ref{fig:exp:all:sets:filtered:hqspre}
and~\ref{fig:exp:all:sets:prepro:hqspre}). The first three places in the
respective rankings of each 
set are taken by solvers based on paradigms~\ref*{paradigm:exp},
\ref*{paradigm:ghostq}, \ref*{paradigm:caqe}, and
\ref*{paradigm:heretic}. That is, solvers \qstsdefnobreaksym,
\depqbfprefixopt, and \quterandom  (paradigms~\ref*{paradigm:qsts}
and~\ref*{paradigm:depqbf}) are not among the three top-performing solvers.

There is a large performance
diversity between different solvers based on the same paradigm. For
example, the expansion solver \dynqbf is ranked last on three sets, which is
in contrast to the overall good performance of the expansion solvers \rareqs
and \revqfun. Likewise, there is a difference  between the
QCDCL solvers \depqbfprefixopt and \quterandom. Such differences between
implementations of the same solving paradigm (or proof system) can be
attributed to the fact that the solvers might apply different heuristics to
explore the search space to find a proof.

The numbers of instances solved uniquely by a particular solver (columns
\emph{U} in Tables~\ref{fig:exp:all:sets:filtered:bloqqer}
to~\ref{fig:exp:all:sets:prepro:hqspre}) highlight the strengths of solvers
such as \qstsdefnobreaksym, \dynqbf, and \depqbfprefixopt which do not show
top performance in the overall rankings. Most notably \dynqbf by far solved
the largest number of instances uniquely on preprocessed sets
$\instancesprefilteredNEW$ (Table~\ref{fig:exp:all:sets:prepro:bloqqer}) and
$\instancesprefilteredHQSpreNEW$
(Table~\ref{fig:exp:all:sets:prepro:hqspre}). With respect to uniquely solved
instances, \qstsdefnobreaksym is second after \dynqbf on set
$\instancesprefilteredNEW$, and \depqbfprefixopt solved the largest number of
instances uniquely on set $\instancesnoprefilteredNEW$
(Table~\ref{fig:exp:all:sets:filtered:bloqqer}).

Towards a more fine-grained analysis of solver performance, we consider the
number of qblocks of instances solved by individual solvers and in
total by solving paradigms. Table~\ref{fig:exp:all:sets:stats} shows related
average and median numbers of qblocks. In general, averages are greater 
for instances from filtered sets ($\instancesnoprefilteredNEW$ and
$\instancesnoprefilteredHQSpreNEW$) than from preprocessed ones
($\instancesprefilteredNEW$ and $\instancesprefilteredHQSpreNEW$), since
preprocessing reduces the numbers of qblocks
(cf.~Figure~\ref{fig:exp:scatter}). The difference in averages between solvers
based on the same paradigm, e.g., \dynqbf and \revqfun in set
$\instancesnoprefilteredNEW$, is due to few solved instances having many
qblocks (up to more than 1000). 

Although the median number of qblocks of instances in \emph{every} considered
set is three (due to alternation bias), the median number of instances solved
by certain solvers as shown in Table~\ref{fig:exp:all:sets:stats} is greater
than three. For example, this is the case for the QCDCL solvers
\depqbfprefixopt and \quterandom on sets $\instancesnoprefilteredNEW$,
$\instancesprefilteredNEW$, and $\instancesnoprefilteredHQSpreNEW$
(\depqbfprefixopt only).  Moreover, QCDCL is the solving paradigm with the
greatest median (6.0 in set $\instancesprefilteredNEW$) among all sets when
considering instances solved by any solver based on a particular paradigm
(rows ``$\bigcup$''). \ijtihad has the greatest median among expansion solvers,
\qstsdefnobreaksym and \heretic have a median of 5.0 on sets
$\instancesnoprefilteredNEW$ and $\instancesprefilteredNEW$, and \caqe has a
median of 5.0 on set $\instancesnoprefilteredNEW$.  These statistics indicate
that there are solvers which tend to perform well on instances with relatively
many qblocks, which however is not reflected in overall
rankings in Tables~\ref{fig:exp:all:sets:filtered:bloqqer}
to~\ref{fig:exp:all:sets:prepro:hqspre} as many of these solvers are \emph{not} among
the top-performing \nolinebreak ones.

\begin{table}[t]
\caption{Instances solved in classes by numbers of
  qblocks (\#q) and numbers of formulas in each class (\#f) for sets
  $\instancesnoprefilteredNEW$
  (\ref{fig:exp:all:sets:filtered:bloqqer:qblocks}),
  $\instancesprefilteredNEW$
  (\ref{fig:exp:all:sets:prepro:bloqqer:qblocks}),
  $\instancesnoprefilteredHQSpreNEW$
  (\ref{fig:exp:all:sets:filtered:hqspre:qblocks}), 
  $\instancesprefilteredHQSpreNEW$
  (\ref{fig:exp:all:sets:prepro:hqspre:qblocks}). Only class winners
  (bold face) are shown, paradigms (\emph{P:}) are indicated in the first~row.}
\addtocounter{table}{-1}
\begin{minipage}[b]{0.48\textwidth}
\begin{center}
\subfloat[Set $\instancesnoprefilteredNEW$ filtered by \bloqqer.]{
{\setlength\tabcolsep{0.15cm}

\begin{tabular}{lrrrrrr}
\hline
\emph{P:} & & \ref*{paradigm:exp} & \ref*{paradigm:ghostq} & \ref*{paradigm:caqe} &
 \ref*{paradigm:heretic} & \ref*{paradigm:depqbf} \\
\#q & \#f 
& \begin{turn}{80}\revqfun\end{turn}
& \begin{turn}{80}\ghostqcegar\end{turn} 
& \begin{turn}{80}\caqe\end{turn} & \begin{turn}{80}\heretic\end{turn}
& \begin{turn}{80}\depqbfprefixopt\end{turn}  \\
\hline
2 & 63 & 17 & \textbf{32}  & 5 & 2 & 6  \\
3 & 215 & \textbf{101} & 89  & 56 & 50 & 47  \\
4--10 & 63 & 25 & 4  & 25 & \textbf{34} & 14  \\
11--20 & 36 & 6 & 3  & 10 & 11 & \textbf{20}  \\
21-- & 60 & 25 & 17  & \textbf{30} & 25 & 28  \\
\hline
2--3 & 278 & 118 & \textbf{121}  & 61 & 52 & 53  \\
4-- & 159 & 56 & 24  & 65 & \textbf{70} & 62 \\
\hline
\end{tabular}

}%end: setlength
\label{fig:exp:all:sets:filtered:bloqqer:qblocks}
}%end subfloat
\end{center}
\end{minipage}
\hfill
\begin{minipage}[b]{0.48\textwidth}
\begin{center}
\subfloat[\mbox{Set $\instancesprefilteredNEW$ preprocessed by \bloqqer.}]{
{\setlength\tabcolsep{0.15cm}

\begin{tabular}{lrrrrr}
\hline
\emph{P:} & & \ref*{paradigm:exp} & \ref*{paradigm:caqe} &
 \ref*{paradigm:heretic} & \ref*{paradigm:exp} \\
\#q & \#f & \begin{turn}{80}\rareqs\end{turn}
& \begin{turn}{80}\caqe\end{turn}
& \begin{turn}{80}\heretic\end{turn} 
& \begin{turn}{80}\dynqbf\end{turn} \\
\hline
2 & 65 & 16 & 15 & 13 &  \textbf{24} \\
3 & 218 & 80 & \textbf{81} & 65  & 18 \\
4--10 & 59 & 37 & 26 & \textbf{38} &  13 \\
11-20 & 53 & 25 & 25 & \textbf{31}  & 4 \\
21-- & 42 & 17 & \textbf{22} & 17 &  6 \\
\hline
2--3 & 283 & \textbf{96} & \textbf{96} & 78 &  42 \\
4-- & 154 & 79 & 73 & \textbf{86} &  23 \\
\hline
\end{tabular}

}%end: setlength
\label{fig:exp:all:sets:prepro:bloqqer:qblocks}
}%end subfloat
\end{center}
\end{minipage}

\medskip

\begin{minipage}[b]{0.48\textwidth}
\begin{center}
\subfloat[Set $\instancesnoprefilteredHQSpreNEW$ filtered by \hqspre.]{
{\setlength\tabcolsep{0.15cm}

\begin{tabular}{lrrrrr}
\hline
\emph{P:} & & \ref*{paradigm:ghostq} & \ref*{paradigm:exp} &
 \ref*{paradigm:caqe} & \ref*{paradigm:depqbf} \\
\#q & \#f & 
\begin{turn}{80}\ghostqcegar\end{turn} & 
\begin{turn}{80}\revqfun\end{turn} & 
\begin{turn}{80}\caqe\end{turn} & 
\begin{turn}{80}\depqbfprefixopt\end{turn} \\ 
  \hline
2 & 70 & \textbf{36} & 18 & 5 & 7  \\
3 & 145 & 62 & \textbf{71} & 33 & 23  \\
4--10 & 26 & 3 & 5 & \textbf{7} & \textbf{7} \\
11--20 & 30 & 3 & 5 & 8 & \textbf{16}  \\
21-- & 41 & 8 & 11 & \textbf{15} & 11  \\
\hline
2--3 & 215 & \textbf{98} & 89 & 38 & 30  \\
4-- & 97& 14 & 21 & 30 & \textbf{34} \\
\hline
\end{tabular}

}%end: setlength
\label{fig:exp:all:sets:filtered:hqspre:qblocks}
}%end subfloat
\end{center}
\end{minipage}
\hfill
\begin{minipage}[b]{0.48\textwidth}
\begin{center}
\subfloat[\mbox{Set $\instancesprefilteredHQSpreNEW$ preprocessed by \hqspre.}]{
{\setlength\tabcolsep{0.15cm}

\begin{tabular}{lrrrrr}
\hline
\emph{P:} & & \ref*{paradigm:caqe} & \ref*{paradigm:heretic} &
 \ref*{paradigm:depqbf} & \ref*{paradigm:exp} \\
\#q & \#f & 
\begin{turn}{80}\caqe\end{turn} & 
\begin{turn}{80}\heretic\end{turn} & 
\begin{turn}{80}\depqbfprefixopt\end{turn} & 
\begin{turn}{80}\dynqbf\end{turn} \\
\hline
2 & 70 & 18 &  15 &  15 &  \textbf{24} \\
3 & 145 & \textbf{67} &  42 &  24  & 14 \\
4--10 & 26 & 6 &  \textbf{10} &  7  & 5 \\
11--20 & 40 & 14 &  15 &  \textbf{20}  & 2 \\
21-- & 31 & \textbf{9} &  5  & 6  & 0 \\
\hline
2--3 & 215 & \textbf{85} &  57  & 39  & 38 \\
4-- & 97 & 29  & 30 &  \textbf{33}  & 7 \\
\hline
\end{tabular}
  
}%end: setlength
\label{fig:exp:all:sets:prepro:hqspre:qblocks}
}%end subfloat
\end{center}
\end{minipage}
\label{fig:exp:all:sets:qblocks}
\refstepcounter{table}
\end{table}

\subsection{Solved Instances: Class-Based Analysis}
\label{sec:class:analysis}

Motivated by the above observations related to median numbers of qblocks of
solved instances, we aim to provide a more detailed picture of the strengths
of the different solvers and implemented solving paradigms. To this end, we
analyze the \emph{numbers of solved instances in classes defined by their numbers of
qblocks}.

Tables~\ref{fig:exp:all:sets:filtered:bloqqer:qblocks}
to~\ref{fig:exp:all:sets:prepro:hqspre:qblocks} show the numbers of
instances that were solved in the individual classes in the considered
sets.  Only class winners are shown (bold face),\footnote{We refer to
  the appendix for complete tables.}  i.e., solvers that solved the
largest number of instances in at least one class, where ties are not
broken. The bottom rows of the tables show statistics
for instances with up to three (row ``2--3'') and more than three
qblocks (row ``4--'').  

The \emph{five different class winners} \revqfun, \ghostqcegar, \caqe,
\heretic, and \depqbfprefixopt in set $\instancesnoprefilteredNEW$ (Table~\ref{fig:exp:all:sets:filtered:bloqqer:qblocks})
implement \emph{five different solving paradigms} (rows \emph{P:}). 
In set $\instancesprefilteredNEW$ (Table~\ref{fig:exp:all:sets:prepro:bloqqer:qblocks})
the four class winners implement three different
paradigms. In sets $\instancesnoprefilteredHQSpreNEW$ and
$\instancesprefilteredHQSpreNEW$ 
(Tables~\ref{fig:exp:all:sets:filtered:hqspre:qblocks}
and~\ref{fig:exp:all:sets:prepro:hqspre:qblocks}), there are 
four different paradigms implemented
in the respective four class winners. Overall, with respect to all four benchmark sets, there
are seven different solvers out of the 11 considered ones that win in a
class. These class winners implement five out
of the six paradigms listed in Section~\ref{sect:setup:experiments},
all except paradigm~\ref*{paradigm:qsts} implemented in
\qstsdefnobreaksym.   

Notably, class winners are not always 
overall top-ranked, and an overall top-ranked 
solver does not always win a class. For example, \rareqs is ranked
third in set $\instancesnoprefilteredNEW$ (Table~\ref{fig:exp:all:sets:filtered:bloqqer}) and second in set
$\instancesprefilteredHQSpreNEW$ (Table~\ref{fig:exp:all:sets:prepro:hqspre}) but does not win a class in the
respective set (Tables~\ref{fig:exp:all:sets:filtered:bloqqer:qblocks} and~\ref{fig:exp:all:sets:prepro:hqspre:qblocks}). As an extreme case, \dynqbf is ranked last on 
sets $\instancesprefilteredNEW$ and
$\instancesprefilteredHQSpreNEW$ (Tables~\ref{fig:exp:all:sets:prepro:bloqqer}
and~\ref{fig:exp:all:sets:prepro:hqspre}) but wins the class of instances with
no more than two qblocks (row ``2'' in Tables~\ref{fig:exp:all:sets:prepro:bloqqer:qblocks} and~\ref{fig:exp:all:sets:prepro:hqspre:qblocks}).

Instances with few qblocks are overrepresented in the benchmark
sets. Alternation bias of this kind in general bears the risk of masking the
strengths of certain solvers on underrepresented instances.  The variety of
class winners and paradigms shown in
Tables~\ref{fig:exp:all:sets:filtered:bloqqer:qblocks}
to~\ref{fig:exp:all:sets:prepro:hqspre:qblocks} is not reflected when only
considering overall solver rankings by total numbers of solved instances in
Tables~\ref{fig:exp:all:sets:filtered:bloqqer}
to~\ref{fig:exp:all:sets:prepro:hqspre}.

The expansion solvers \revqfun and \rareqs (paradigm~\ref*{paradigm:exp}) tend to
perform better on instances with relatively few qblocks, while solvers
applying QCDCL (paradigms~\ref*{paradigm:depqbf} and~\ref*{paradigm:heretic}) tend to
perform better on many qblocks. For example, either \depqbfprefixopt
or \heretic win on instances with four or more qblocks (row ``4--'')
in any set. These
statistics are interesting in the context of QBF proof complexity as the
proof systems underlying expansion and QCDCL are 
orthogonal~\cite{beyersdorff_et_al:LIPIcs:2015:4905,DBLP:journals/tcs/JanotaM15}.
\caqe based on paradigm~\ref*{paradigm:caqe} wins on 
instances with 21 or more qblocks (rows ``21--'') in all sets
(Tables~\ref{fig:exp:all:sets:filtered:bloqqer:qblocks}
to~\ref{fig:exp:all:sets:prepro:hqspre:qblocks}). Further, it also
wins on instances with no more than three qblocks in
set~$\instancesprefilteredHQSpreNEW$
(Table~\ref{fig:exp:all:sets:prepro:hqspre:qblocks}). The proof systems
underlying paradigms~\ref*{paradigm:caqe} and~\ref*{paradigm:exp} (expansion) are
orthogonal~\cite{DBLP:conf/cav/Tentrup17}. 
The performance diversity of orthogonal proof systems on instances with
different numbers of qblocks is not reflected in overall rankings
and motivates further, theoretical study in QBF \nolinebreak proof \nolinebreak complexity.

Due to alternation bias, classes of instances
with few qblocks are larger than those with many
qblocks. Hence solvers often win in a class of instances with
many qblocks by only a small margin. For example, the
top-ranked solvers on classes ``4--10'', ``11--20'', and ``21--'' 
tend to be close to each other in terms of solved instances (cf.~appendix). Moreover, solvers
implementing the same paradigm might show diverse performance due to different
heuristics in proof search. 
To consider these factors, 
we carry out a
\emph{class-based analysis of solving paradigms}. To this end,
we count instances solved by any solver implementing a
particular paradigm. This study is related to statistics in rows
``$\bigcup$'' of Table~ \ref{fig:exp:all:sets:stats}.

Tables~\ref{fig:2017:exp:bloqqer37:noprepro:paradigms:solved:qblocks:full:cobra:MAIN}
to~\ref{fig:2017:exp:hqspre:prepro:paradigms:solved:qblocks:full:cobra:MAIN}
show instances solved by each of the solving
paradigms~\ref*{paradigm:exp} to~\ref*{paradigm:heretic} (first row)
in classes of instances. Class winners are highlighted in bold face. 
Paradigm~\ref*{paradigm:exp} (expansion) dominates the other
paradigms on complete benchmark sets (row ``2--''). On instances
obtained by \bloqqer
(Tables~\ref{fig:2017:exp:bloqqer37:noprepro:paradigms:solved:qblocks:full:cobra:MAIN}
and~\ref{fig:2017:exp:bloqqer37:prepro:paradigms:solved:qblocks:full:cobra:MAIN}),
in total only four classes are won by paradigms other than expansion:
class ``2'' by paradigm~\ref*{paradigm:ghostq} (QDPLL) on set
$\instancesnoprefilteredNEW$, class ``11--20'' by
paradigm~\ref*{paradigm:depqbf} (QCDCL) on sets $\instancesnoprefilteredNEW$
and $\instancesprefilteredNEW$, and class ``21--'' by
paradigm~\ref*{paradigm:caqe} (clause selection/abstraction) on set
$\instancesprefilteredNEW$. Regarding the dominance of
paradigm~\ref*{paradigm:exp} (expansion) in
Tables~\ref{fig:2017:exp:bloqqer37:noprepro:paradigms:solved:qblocks:full:cobra:MAIN}
and~\ref{fig:2017:exp:bloqqer37:prepro:paradigms:solved:qblocks:full:cobra:MAIN}, 
we note that four solvers among the considered ones are based on expansion,
while there are at most two solvers implementing the other paradigms.

Performance is more diverse on instances filtered and preprocessed by
\hqspre
(Tables~\ref{fig:2017:exp:hqspre:noprepro:paradigms:solved:qblocks:full:cobra:MAIN}
and~\ref{fig:2017:exp:hqspre:prepro:paradigms:solved:qblocks:full:cobra:MAIN}). There,
paradigms other than expansion either win or are on par with expansion
in nine classes in total. Notably, paradigms~\ref*{paradigm:caqe}
 and~\ref*{paradigm:depqbf}  win in
classes ``4--'' of sets $\instancesprefilteredHQSpreNEW$ and
$\instancesnoprefilteredHQSpreNEW$ containing instances with many
qblocks. Although \caqe (paradigm~\ref*{paradigm:caqe}) is overall
top-ranked on set $\instancesprefilteredHQSpreNEW$
(Table~\ref{fig:exp:all:sets:prepro:hqspre}), the strong performance
of paradigms~\ref*{paradigm:caqe} and~\ref*{paradigm:depqbf} on
instances with many qblocks is not reflected in overall rankings
(Tables~\ref{fig:exp:all:sets:filtered:hqspre}
and~\ref{fig:exp:all:sets:prepro:hqspre}).

%%%%%%%%%%%%%%%%%%%%%%%%%%%%%%%%%%%%%%%%%%%%%

\begin{table}[t]
\caption{Instances solved by solving
  paradigms~\ref*{paradigm:exp} to~\ref*{paradigm:heretic} (cf.~Section~\ref{sect:setup:experiments}) in classes by numbers of
  qblocks (\#q) for sets
  $\instancesnoprefilteredNEW$
  (\ref{fig:2017:exp:bloqqer37:noprepro:paradigms:solved:qblocks:full:cobra:MAIN}),
  $\instancesprefilteredNEW$
  (\ref{fig:2017:exp:bloqqer37:prepro:paradigms:solved:qblocks:full:cobra:MAIN}),
  $\instancesnoprefilteredHQSpreNEW$
  (\ref{fig:2017:exp:hqspre:noprepro:paradigms:solved:qblocks:full:cobra:MAIN}),
  and $\instancesprefilteredHQSpreNEW$~(\ref{fig:2017:exp:hqspre:prepro:paradigms:solved:qblocks:full:cobra:MAIN}).}
\addtocounter{table}{-1}
\begin{minipage}[b]{0.50\textwidth}
\begin{center}
\subfloat[Set $\instancesnoprefilteredNEW$ filtered by \bloqqer.]{
{\setlength\tabcolsep{0.15cm}

\begin{tabular}{lrrrrrr}
\hline
\#q  & \ref*{paradigm:exp} & \ref*{paradigm:ghostq} & \ref*{paradigm:qsts} & \ref*{paradigm:caqe} & \ref*{paradigm:depqbf} & \ref*{paradigm:heretic} \\
\hline
2 &  26 & \textbf{32} & 8 & 6 & 7 & 2 \\
3 &  \textbf{121} & 89 & 43 & 61 & 66 & 50 \\
4--10  & \textbf{38} & 4 & 21 & 27 & 16 & 34 \\
11--20  & 10 & 3 & 8 & 10 & \textbf{20} & 11 \\
21--  & \textbf{33} & 17 & 23 & 30 & 28 & 25 \\
\hline
2--3  & \textbf{147} & 121 & 51 & 67 & 73 & 52 \\
4--  & \textbf{81} & 24 & 52 & 67 & 64 & 70 \\
\hline
2--  & \textbf{228} & 145 & 103 & 134 & 137 & 122 \\
\hline
\end{tabular}

}%end: setlength
\label{fig:2017:exp:bloqqer37:noprepro:paradigms:solved:qblocks:full:cobra:MAIN}
}%end subfloat
\end{center}
\end{minipage}
\hfill
\begin{minipage}[b]{0.50\textwidth}
\begin{center}
\subfloat[\mbox{Set $\instancesprefilteredNEW$ preprocessed by \bloqqer.}]{
{\setlength\tabcolsep{0.15cm}

\begin{tabular}{lrrrrrr}
\hline
\#q  & \ref*{paradigm:exp} & \ref*{paradigm:ghostq} & \ref*{paradigm:qsts} & \ref*{paradigm:caqe} & \ref*{paradigm:depqbf} & \ref*{paradigm:heretic} \\
\hline
2  & \textbf{37} & 3 & 11 & 17 & 10 & 13 \\
3  & \textbf{103} & 53 & 46 & 86 & 40 & 65 \\
4--10  & \textbf{49} & 5 & 25 & 28 & 18 & 38 \\
11--20  & 31 & 9 & 24 & 27 & \textbf{32} & 31 \\
21--  & 18 & 12 & 21 & \textbf{24} & 17 & 17 \\
\hline
2--3  & \textbf{140} & 56 & 57 & 103 & 50 & 78 \\
4--  & \textbf{98} & 26 & 70 & 79 & 67 & 86 \\
\hline
2--  & \textbf{238} & 82 & 127 & 182 & 117 & 164 \\
\hline
\end{tabular}

}%end: setlength
\label{fig:2017:exp:bloqqer37:prepro:paradigms:solved:qblocks:full:cobra:MAIN}
}%end subfloat
\end{center}
\end{minipage}

\medskip

\begin{minipage}[b]{0.50\textwidth}
\begin{center}
\subfloat[Set $\instancesnoprefilteredHQSpreNEW$ filtered by \hqspre.]{
{\setlength\tabcolsep{0.15cm}

\begin{tabular}{lrrrrrr}
\hline
\#q  & \ref*{paradigm:exp} & \ref*{paradigm:ghostq} & \ref*{paradigm:qsts} & \ref*{paradigm:caqe} & \ref*{paradigm:depqbf} & \ref*{paradigm:heretic} \\
\hline
2  & 28 & \textbf{36} & 9 & 6 & 8 & 2 \\
3  & \textbf{85} & 62 & 27 & 36 & 40 & 23 \\
4--10  & \textbf{9} & 3 & 1 & \textbf{9} & 8 & 5 \\
11-20  & 8 & 3 & 7 & 8 & \textbf{16} & 9 \\
21--  & \textbf{15} & 8 & 12 & \textbf{15} & 11 & 10 \\
\hline
2--3  & \textbf{113} & 98 & 36 & 42 & 48 & 25 \\
4--  & 32 & 14 & 20 & 32 & \textbf{35} & 24 \\
\hline
2--  & \textbf{145} & 112 & 56 & 74 & 83 & 49 \\
\hline
\end{tabular}

}%end: setlength
\label{fig:2017:exp:hqspre:noprepro:paradigms:solved:qblocks:full:cobra:MAIN}
}%end subfloat
\end{center}
\end{minipage}
\hfill
\begin{minipage}[b]{0.50\textwidth}
\begin{center}
\subfloat[\mbox{Set $\instancesprefilteredHQSpreNEW$ preprocessed by \hqspre.}]{
{\setlength\tabcolsep{0.15cm}

\begin{tabular}{lrrrrrr}
\hline
\#q  & \ref*{paradigm:exp} & \ref*{paradigm:ghostq} & \ref*{paradigm:qsts} & \ref*{paradigm:caqe} & \ref*{paradigm:depqbf} & \ref*{paradigm:heretic} \\
\hline
2  & \textbf{37} & 7 & 17 & 18 & 21 & 15 \\
3  & \textbf{78} & 40 & 35 & 71 & 40 & 42 \\
4--10  & 10 & 1 & 2 & \textbf{13} & 7 & 10 \\
11--20  & 17 & 6 & 13 & 15 & \textbf{21} & 15 \\
21--  & 8 & 4 & 5 & \textbf{10} & 8 & 5 \\
\hline
2--3  & \textbf{115} & 47 & 52 & 89 & 61 & 57 \\
4--  & 35 & 11 & 20 & \textbf{38} & 36 & 30 \\
\hline
2--  & \textbf{150} & 58 & 72 & 127 & 97 & 87 \\
\hline
\end{tabular}
  
}%end: setlength
\label{fig:2017:exp:hqspre:prepro:paradigms:solved:qblocks:full:cobra:MAIN}
}%end subfloat
\end{center}
\end{minipage}
\label{fig:2017:exp:paradigms:solved:qblocks:full:cobra:MAIN}
\refstepcounter{table}
\end{table}

%%%%%%%%%%%%%%%%%%%%%%%%%%%%%%%%%%%%%%%%%%%%%

\subsection{Virtual Best Solver Analysis}

%%%%%%%%%%%%%%%%%%%%%%%%%%%%%%%%%%%%%%%%%%%%%

\begin{table}[t]
\caption{Instances solved by the
  virtual best solver (VBS) in classes by number of qblocks (\#q), number of
  formulas (\#f) in each class, and relative contribution (\%) of each solver to
  instances solved by the VBS for sets 
  $\instancesnoprefilteredHQSpreNEW$
  (\ref{fig:2017:exp:all:in:one:qblocks:vbs:cobra:noprepro:hqspre}) 
  and $\instancesprefilteredHQSpreNEW$~(\ref{fig:2017:exp:all:in:one:qblocks:vbs:cobra:prepro:hqspre}).}
\addtocounter{table}{-1}
{\scriptsize

%%%%%%%%%

\subfloat[Set $\instancesnoprefilteredHQSpreNEW$ filtered by \hqspre.]{
{\setlength\tabcolsep{0.15cm}
\begin{tabular}{lrrrrrrrrrrrrr}
\hline
\#q & \#f & \begin{turn}{80}VBS\end{turn} & 
\begin{turn}{80}\ghostqcegar\end{turn} & 
\begin{turn}{80}\revqfun\end{turn} & 
\begin{turn}{80}\caqe\end{turn} & 
\begin{turn}{80}\depqbfprefixopt\end{turn} & 
\begin{turn}{80}\qstsdefnobreaksym\end{turn} & 
\begin{turn}{80}\rareqs\end{turn} & 
\begin{turn}{80}\heretic\end{turn} & 
\begin{turn}{80}\quterandom\end{turn} & 
\begin{turn}{80}\dynqbf\end{turn} &
\begin{turn}{80}\qesto\end{turn} & 
\begin{turn}{80}\ijtihad\end{turn} \\ 
\hline

2 & 70 & 46 & \textbf{41.3} & 6.5 & 6.5 & 6.5 & 6.5 & 0.0 & 0.0 & 0.0 & 30.4 & 2.1 & 0.0  \\
3 & 145 & 89 & 12.3 & \textbf{33.7} & 2.2 & 2.2 & 15.7 & 22.4 & 0.0 & 3.3 & 2.2 & 4.4 & 1.1  \\
4--10 & 26 & 19 & 5.2 & 0.0 & \textbf{26.3} & \textbf{26.3} & 0.0 & 0.0 & 0.0 & 15.7 & 10.5 & 10.5 & 5.2  \\
11-20 & 30 & 18 & 0.0 & 0.0 & 11.1 & \textbf{50.0} & 27.7 & 5.5 & 0.0 & 0.0 & 5.5 & 0.0 & 0.0  \\
21-- & 41 & 21 & 4.7 & 14.2 & 19.0 & 9.5 & \textbf{28.5} & 14.2 & 0.0 & 0.0 & 9.5 & 0.0 & 0.0  \\
\hline
2--3 & 215 & 135 & 22.2 & \textbf{24.4} & 3.7 & 3.7 & 12.5 & 14.8 & 0.0 & 2.2 & 11.8 & 3.7 & 0.7  \\
4-- & 97 & 58 & 3.4 & 5.1 & 18.9 & \textbf{27.5} & 18.9 & 6.8 & 0.0 & 5.1 & 8.6 & 3.4 & 1.7  \\
\hline
2-- & 312 & 193 & 16.5 & \textbf{18.6} & 8.2 & 10.8 & 14.5 & 12.4 & 0.0 & 3.1 & 10.8 & 3.6 & 1.0  \\
\hline
\end{tabular}
}%end: setlength
\label{fig:2017:exp:all:in:one:qblocks:vbs:cobra:noprepro:hqspre}
}%end subfloat

%%%%%%%%%

\subfloat[Set $\instancesprefilteredHQSpreNEW$ preprocessed by \hqspre.]{
{\setlength\tabcolsep{0.15cm}
\begin{tabular}{lrrrrrrrrrrrrr}
\hline
\#q & \#f & \begin{turn}{80}VBS\end{turn} & 
\begin{turn}{80}\caqe\end{turn} & 
\begin{turn}{80}\rareqs\end{turn} &
\begin{turn}{80}\qesto\end{turn} & 
\begin{turn}{80}\revqfun\end{turn} &  
\begin{turn}{80}\heretic\end{turn} & 
\begin{turn}{80}\qstsdefnobreaksym\end{turn} & 
\begin{turn}{80}\depqbfprefixopt\end{turn} & 
\begin{turn}{80}\quterandom\end{turn} & 
\begin{turn}{80}\ijtihad\end{turn} & 
\begin{turn}{80}\ghostqcegar\end{turn} & 
\begin{turn}{80}\dynqbf\end{turn} \\
\hline
2 & 70 & 40 & 7.5 & 17.5 & 2.5 & 7.5 & 2.5 & 10.0 & 10.0 & 0.0 & 0.0 & 2.5 & \textbf{40.0}  \\
3 & 145 & 87 & 9.1 & \textbf{40.2} & 8.0 & 12.6 & 1.1 & 6.8 & 0.0 & 8.0 & 3.4 & 4.5 & 5.7  \\
4--10 & 26 & 20 & \textbf{25.0} & 10.0 & 15.0 & 5.0 & 0.0 & 0.0 & \textbf{25.0} & 5.0 & 5.0 & 0.0 & 10.0  \\
11-20 & 40 & 26 & 3.8 & 19.2 & 7.6 & 0.0 & 7.6 & 26.9 & \textbf{30.7} & 0.0 & 0.0 & 0.0 & 3.8  \\
21-- & 31 & 11 & 9.0 & \textbf{27.2} & 9.0 & 9.0 & 0.0 & \textbf{27.2} & 9.0 & 9.0 & 0.0 & 0.0 & 0.0  \\
\hline
2--3 & 215 & 127 & 8.6 & \textbf{33.0} & 6.2 & 11.0 & 1.5 & 7.8 & 3.1 & 5.5 & 2.3 & 3.9 & 16.5  \\
4-- & 97 & 57 & 12.2 & 17.5 & 10.5 & 3.5 & 3.5 & 17.5 & \textbf{24.5} & 3.5 & 1.7 & 0.0 & 5.2  \\
\hline
2-- & 312 & 184 & 9.7 & \textbf{28.2} & 7.6 & 8.6 & 2.1 & 10.8 & 9.7 & 4.8 & 2.1 & 2.7 & 13.0  \\
\hline
\end{tabular}

}%end: setlength
\label{fig:2017:exp:all:in:one:qblocks:vbs:cobra:prepro:hqspre}
}%end subfloat

}%end:scriptsize
\label{fig:2017:exp:all:in:one:qblocks:vbs:cobra}
\refstepcounter{table}
\end{table}

We strengthen our above observations of performance diversity of
solvers and solving paradigms with respect to numbers of qblocks by a
\emph{virtual best solver (VBS)} analysis, which is common in
QBF~\cite{DBLP:journals/fuin/MarinNPTG16} and SAT
competitions~(cf.~\cite{DBLP:journals/ai/BalyoBIS16}).
The VBS is an
ideal portfolio where the solving time of the fastest solver on an 
instance is attributed to the VBS. Thus the VBS reflects the best
performance that can be achieved when running a set of solvers in
parallel on an instance.

Tables~\ref{fig:2017:exp:all:in:one:qblocks:vbs:cobra:noprepro:hqspre}
and~\ref{fig:2017:exp:all:in:one:qblocks:vbs:cobra:prepro:hqspre} show
numbers of instances solved by the VBS in classes for sets
$\instancesnoprefilteredHQSpreNEW$ and
$\instancesprefilteredHQSpreNEW$ and the relative contribution of 
solvers (percentage) to the VBS in terms of solved
instances.  Similar to instances solved in classes
(Tables~\ref{fig:exp:all:sets:filtered:bloqqer:qblocks}
to~\ref{fig:exp:all:sets:prepro:hqspre:qblocks}), the VBS contributions
differ and provide a more
fine-grained picture of the strengths of solvers and solving paradigms
than the VBS contributions on the entire benchmark set (rows ``2--''
in
Tables~\ref{fig:2017:exp:all:in:one:qblocks:vbs:cobra:noprepro:hqspre}
and~\ref{fig:2017:exp:all:in:one:qblocks:vbs:cobra:prepro:hqspre}). In
the following, we comment on general VBS statistics for all considered benchmark sets, with a focus on
sets~$\instancesnoprefilteredHQSpreNEW$ and
$\instancesprefilteredHQSpreNEW$ generated using \hqspre. We refer to the
appendix for tables  
related to sets~$\instancesnoprefilteredNEW$ and
$\instancesprefilteredNEW$ generated using \nolinebreak \bloqqer.

On all benchmark sets the VBS solved between 50\% and 70\% more
instances than the single overall best solver
(Tables~\ref{fig:exp:all:sets:filtered:bloqqer}
to~\ref{fig:exp:all:sets:prepro:hqspre}). These results highlight the
complementary strengths of solvers and solving paradigms that are not
among the top-ranked ones.  On
each of the four benchmark sets, there are five different solvers,
respectively, which have the largest VBS contribution in a
class. Interestingly, from the respective overall winning solvers
(Tables~\ref{fig:exp:all:sets:filtered:bloqqer}
to~\ref{fig:exp:all:sets:prepro:hqspre}), only \rareqs on set
$\instancesprefilteredNEW$ also has the largest VBS contribution on
the entire benchmark set. While \rareqs is ranked second
on set $\instancesprefilteredHQSpreNEW$
(Table~\ref{fig:exp:all:sets:prepro:hqspre}), it has the largest
overall VBS contribution (row ``2--'' in
Table~\ref{fig:2017:exp:all:in:one:qblocks:vbs:cobra:prepro:hqspre}).

Consistent with Tables~\ref{fig:exp:all:sets:prepro:bloqqer:qblocks}
and~\ref{fig:exp:all:sets:prepro:hqspre:qblocks}, where \dynqbf solved
the largest number of instances in class ``2'' of sets
$\instancesprefilteredNEW$ and $\instancesprefilteredHQSpreNEW$, it
has the largest VBS contributions 
in this class 
(cf.~Table~\ref{fig:2017:exp:all:in:one:qblocks:vbs:cobra:prepro:hqspre} and appendix)
although it is ranked last in overall rankings 
(Tables~\ref{fig:exp:all:sets:prepro:bloqqer}
and~\ref{fig:exp:all:sets:prepro:hqspre}). The large VBS contributions of
\dynqbf conform to the fact that it solved the largest numbers of
instances uniquely in sets $\instancesprefilteredNEW$ and
$\instancesprefilteredHQSpreNEW$. Similar observations regarding VBS
contributions of solvers that are not top-ranked were made in the
context of SAT solver competitions~\cite{DBLP:conf/sat/XuHHL12}.

\qstsdefnobreaksym neither is among the overall top-ranked
solvers (Tables~\ref{fig:exp:all:sets:filtered:bloqqer}
to~\ref{fig:exp:all:sets:prepro:hqspre}) nor among the class winners 
(Tables~\ref{fig:exp:all:sets:filtered:bloqqer:qblocks}
to~\ref{fig:exp:all:sets:prepro:hqspre:qblocks}), yet it has the largest
VBS contribution in class ``21--'' on all sets except
$\instancesprefilteredHQSpreNEW$
(Table~\ref{fig:2017:exp:all:in:one:qblocks:vbs:cobra:prepro:hqspre}),
where it is on \nolinebreak par \nolinebreak with \nolinebreak \rareqs.

Similar to the analysis presented in 
Tables~\ref{fig:2017:exp:bloqqer37:noprepro:paradigms:solved:qblocks:full:cobra:MAIN}
to~\ref{fig:2017:exp:hqspre:prepro:paradigms:solved:qblocks:full:cobra:MAIN},
we analyze the \emph{VBS contribution of each solving paradigm} for
sets $\instancesnoprefilteredHQSpreNEW$ and
$\instancesprefilteredHQSpreNEW$ in
Tables~\ref{fig:2017:exp:hqspre:noprepro:all:in:one:paradigms:vbs:qblocks:full:cobra}
and~\ref{fig:2017:exp:hqspre:prepro:all:in:one:paradigms:vbs:qblocks:full:cobra},
respectively. We refer to 
the appendix for tables related to sets
$\instancesnoprefilteredNEW$ and $\instancesprefilteredNEW$. Considering instances with many qblocks (row ``4--''),
paradigm~\ref*{paradigm:depqbf} (QCDCL) has the largest contribution
in set $\instancesnoprefilteredHQSpreNEW$ and is on par with
paradigm~\ref*{paradigm:exp} (expansion) in set
$\instancesprefilteredHQSpreNEW$. This is remarkable, given that
paradigm~\ref*{paradigm:exp}, where four solvers are based on, clearly
has the largest VBS contribution on the entire sets (rows
``2--''). However, only two solvers implement
paradigm~\ref*{paradigm:depqbf}.

\begin{table}[t]
\caption{Instances solved by the
  virtual best solver (VBS) in classes by number of qblocks (\#q), number of
  formulas (\#f) in each class, and relative contribution (\%) of solving paradigms to
  instances solved by the VBS for sets
  $\instancesnoprefilteredHQSpreNEW$
  (\ref{fig:2017:exp:hqspre:noprepro:all:in:one:paradigms:vbs:qblocks:full:cobra}),
  and $\instancesprefilteredHQSpreNEW$~(\ref{fig:2017:exp:hqspre:prepro:all:in:one:paradigms:vbs:qblocks:full:cobra}).}
\addtocounter{table}{-1}

\begin{center}

{

\subfloat[Set $\instancesnoprefilteredHQSpreNEW$ filtered by \hqspre.]{
{\setlength\tabcolsep{0.15cm}
\begin{tabular}{lrrrrrrrr}
\hline
\#q & \#f & VBS & \ref*{paradigm:exp} & \ref*{paradigm:ghostq} & \ref*{paradigm:qsts} & \ref*{paradigm:caqe} & \ref*{paradigm:depqbf} & \ref*{paradigm:heretic} \\
\hline
2 & 70 & 46 & 36.9 & \textbf{41.3} & 6.5 & 8.6 & 6.5 & 0.0  \\
3 & 145 & 89 & \textbf{59.5} & 12.3 & 15.7 & 6.7 & 5.6 & 0.0  \\
4--10 & 26 & 19 & 15.7 & 5.2 & 0.0 & 36.8 & \textbf{42.1} & 0.0  \\
11--20 & 30 & 18 & 11.1 & 0.0 & 27.7 & 11.1 & \textbf{50.0} & 0.0  \\
21-- & 41 & 21 & \textbf{38.0} & 4.7 & 28.5 & 19.0 & 9.5 & 0.0  \\
\hline
2--3 & 215 & 135 & \textbf{51.8} & 22.2 & 12.5 & 7.4 & 5.9 & 0.0  \\
4-- & 97 & 58 & 22.4 & 3.4 & 18.9 & 22.4 & \textbf{32.7} & 0.0  \\
\hline
2-- & 312 & 193 & \textbf{43.0} & 16.5 & 14.5 & 11.9 & 13.9 & 0.0  \\
\hline
\end{tabular}
}%end: setlength
\label{fig:2017:exp:hqspre:noprepro:all:in:one:paradigms:vbs:qblocks:full:cobra}
}%end subfloat

%%%%%%%%%

\subfloat[Set $\instancesprefilteredHQSpreNEW$ preprocessed by \hqspre.]{
{\setlength\tabcolsep{0.15cm}
\begin{tabular}{lrrrrrrrr}
\hline
\#q & \#f & VBS & \ref*{paradigm:exp} & \ref*{paradigm:ghostq} & \ref*{paradigm:qsts} & \ref*{paradigm:caqe} & \ref*{paradigm:depqbf} & \ref*{paradigm:heretic} \\
\hline
2 & 70 & 40 & \textbf{65.0} & 2.5 & 10.0 & 10.0 & 10.0 & 2.5  \\
3 & 145 & 87 & \textbf{62.0} & 4.5 & 6.8 & 17.2 & 8.0 & 1.1  \\
4--10 & 26 & 20 & 30.0 & 0.0 & 0.0 & \textbf{40.0} & 30.0 & 0.0  \\
11--20 & 40 & 26 & 23.0 & 0.0 & 26.9 & 11.5 & \textbf{30.7} & 7.6  \\
21-- & 31 & 11 & \textbf{36.3} & 0.0 & 27.2 & 18.1 & 18.1 & 0.0  \\
\hline
2--3 & 215 & 127 & \textbf{62.9} & 3.9 & 7.8 & 14.9 & 8.6 & 1.5  \\
4-- & 97 & 57 & \textbf{28.0} & 0.0 & 17.5 & 22.8 & \textbf{28.0} & 3.5  \\
\hline
2-- & 312 & 184 & \textbf{52.1} & 2.7 & 10.8 & 17.3 & 14.6 & 2.1  \\
\hline
\end{tabular}
}%end: setlength
\label{fig:2017:exp:hqspre:prepro:all:in:one:paradigms:vbs:qblocks:full:cobra}
}%end subfloat

%%%%%%%%%

}%end: font size

\end{center}

\label{fig:2017:exp:all:in:one:paradigms:vbs:qblocks:full:cobra}
\refstepcounter{table}
\end{table}

%%%%%%%%%%%%%%%%%%%%%%%%%%%%%%%%%%%%%%%%%%%%%%%%%%%%%%%%%%%%%%

\subsection{Discussion}

In the following, we discuss threats to the validity of our study and related issues.

\paragraph{\textbf{Heuristics.}}  
The performance of solvers implementing the same paradigm might be
diverse due to different heuristics applied in proof search. To
comprehensively evaluate the impact of heuristics, it is necessary to
consider further syntactic parameters of instances other than
alternations, such as ratio of variables per clause, size of clauses,
and the like. In our study, we focused on alternations as they impact
the theoretical hardness of PCNFs, thus resulting in a larger
complexity landscape than, e.g., in propositional logic (SAT).  To
even out the effects of heuristics, we studied and observed
performance diversity of \emph{paradigms}
(Tables~\ref{fig:2017:exp:paradigms:solved:qblocks:full:cobra:MAIN}
and~\ref{fig:2017:exp:all:in:one:paradigms:vbs:qblocks:full:cobra}). Such
diversity cannot be explained by different heuristics, in contrast to
diversity between individual solvers based on the same \nolinebreak
paradigm.

\paragraph{\textbf{Dominance of single solvers and paradigms.}}
We are not aware of solvers being specifically targeted to instances
with a particular number of alternations.  Similar to the effects of
heuristics, we even out a potential dominance of single solvers and
overrepresented paradigms in solvers by a paradigm-based analysis
(Tables~\ref{fig:2017:exp:paradigms:solved:qblocks:full:cobra:MAIN}
and~\ref{fig:2017:exp:all:in:one:paradigms:vbs:qblocks:full:cobra}). This
provides a more comprehensive picture of the strengths of different
paradigms. This way, e.g., we observed remarkable results regarding the VBS
contribution of QCDCL on instances with many alternations
(Table~\ref{fig:2017:exp:all:in:one:paradigms:vbs:qblocks:full:cobra}).

\paragraph{\textbf{Choice of benchmarks and solvers.}} 
The benchmarks we considered 
contain few instances with many alternations, which follows from
alternation bias in original benchmarks
(cf.~Section~\ref{sect:setup:experiments}). We observed
performance diversity in the large classes ``2--3'' and ``4--'', which
is more robust than in smaller classes containing fewer
instances. Class ``4--'' is the largest one with many alternations
that can be selected in the given benchmarks.  Our choice of
solvers was predetermined by the ranking of the top-performing solvers in the PCNF track of
\mbox{QBFEVAL'17}. 

\paragraph{\textbf{Relation to QBF proof complexity.}}
We emphasize that our study does \emph{not} show that certain proof systems
\emph{provably} perform differently with respect to alternations. This is an
open research problem in QBF proof complexity.

\paragraph{\textbf{Overrepresented problems and different prenex forms.}}
Several QBF encodings of a problem with different numbers of
alternations may exist. Hence in the instance classes we defined by
alternations certain problems might be overrepresented. These problems
may be detected based on detailed information about the encoding
process. However, such information is often not available for PCNF
benchmarks. A related issue is the impact of different quantifier prefixes in PCNFs on solver performance, which was studied in
theory~\cite{DBLP:conf/aaai/BeyersdorffCJ16}
and practice~\cite{DBLP:conf/sat/EglySTWZ03}.

%%%%%%%%%%%%%%%%%%%%%%%%%%%%%%%%%%%%%%%%%%%%%%%%%%%%%%%%%%%%%%%%%%%%%%%%%%%%%%%%
%%%%%%%%%%%%%%%%%%%%%%%%%%%%%%%%%%%%%%%%%%%%%%%%%%%%%%%%%%%%%%%%%%%%%%%%%%%%%%%%

\section{Conclusion}

We analyzed the effects of quantifier alternations on the evaluation of QBF
solvers. Our empirical results indicate that the performance of solvers based on
\emph{different solving paradigms} substantially varies on classes of formulas
defined by their numbers of alternations. While the \emph{theoretical
hardness} of QBFs in prenex CNF with a particular number of alternations is
naturally related to levels in the polynomial hierarchy, our study \emph{a
posteriori} sheds light on solver performance \emph{observed in practice}. We
observed a substantial performance diversity of solvers based on orthogonal QBF proof
systems~\cite{beyersdorff_et_al:LIPIcs:2015:4905,DBLP:journals/tcs/JanotaM15,DBLP:conf/cav/Tentrup17} 
on instances with different numbers of alternations, e.g., expansion and Q-resolution. Thereby, our work is in
line with a recent focus on alternations in QBF proof
complexity~\cite{DBLP:conf/innovations/BeyersdorffBH18,DBLP:conf/fsttcs/BeyersdorffHP17,DBLP:journals/toct/Chen17}. 
As a future direction in practice, and motivated by virtual best solver statistics we presented, it is promising to combine orthogonal approaches to 
leverage their individual strengths in a single QBF solver.

The class- and paradigm-based performance analysis we presented is a methodology to evaluate QBF
solvers that takes quantifier alternations of under- and
overrepresented instances into account. This is necessary to highlight
the strengths of solving paradigms in a comprehensive way.
In doing so, we aim to reach out to users of QBF technology who
are inexperienced with solver implementations and look for solvers
that are suitable to solve a particular problem.  Ultimately, QBF
technology must be improved as a general approach to tackle PSPACE \nolinebreak problems.

%%%%%%%%%%%%%%%%%%%%%%%%%%%%%%%%%%%%%%%%%%%%%%%%%%%%%%%%%%%%%%%%%%%%%%%%%%%%%%%%
%%%%%%%%%%%%%%%%%%%%%%%%%%%%%%%%%%%%%%%%%%%%%%%%%%%%%%%%%%%%%%%%%%%%%%%%%%%%%%%%

%%%%%%%%%%%%%%%%%%%%%%%%%%%%%%%%%%%%%%%%%%%%%%%%%%%%%%%%%%%%%%%%%%%%%%%%%%%%%%%%
%%%%%%%%%%%%%%%%%%%%%%%%%%%%%%%%%%%%%%%%%%%%%%%%%%%%%%%%%%%%%%%%%%%%%%%%%%%%%%%%

\clearpage

\begin{appendix}

\section{Appendix}

\subsection{Additional Experimental Data}

\begin{table}[ht]
\caption{ Instances solved in set
$\instancesnoprefilteredNEW$ filtered by \bloqqer with respect to classes by
number of qblocks (\#q) and number of formulas in each class (\#f).}
\label{fig:2017:exp:bloqqer37:noprepro:solved:qblocks:full:cobra}
\begin{center}
{\setlength\tabcolsep{0.15cm}
\begin{tabular}{lrrrrrrrrrrrr}
\hline
\#q & \#f 
& \begin{turn}{80}\revqfun\end{turn}
& \begin{turn}{80}\ghostqcegar\end{turn} & \begin{turn}{80}\rareqs\end{turn}
& \begin{turn}{80}\caqe\end{turn} & \begin{turn}{80}\heretic\end{turn}
& \begin{turn}{80}\depqbfprefixopt\end{turn} 
& \begin{turn}{80}\ijtihad\end{turn} 
& \begin{turn}{80}\qstsdefnobreaksym\end{turn}
& \begin{turn}{80}\quterandom\end{turn} & \begin{turn}{80}\qesto\end{turn}
& \begin{turn}{80}\dynqbf\end{turn} \\
\hline
2 & 63 & 17 & \textbf{32} & 2 & 5 & 2 & 6 & 2 & 8 & 2 & 4 & 18 \\
3 & 215 & \textbf{101} & 89 & 62 & 56 & 50 & 47 & 49 & 43 & 36 & 35 & 19 \\
4--10 & 63 & 25 & 4 & 33 & 25 & \textbf{34} & 14 & 32 & 21 & 10 & 6 & 4 \\
11--20 & 36 & 6 & 3 & 6 & 10 & 11 & \textbf{20} & 4 & 8 & 7 & 8 & 1 \\
21-- & 60 & 25 & 17 & 23 & \textbf{30} & 25 & 28 & 23 & 23 & 22 & 23 & 5 \\
\hline
2--3 & 278 & 118 & \textbf{121} & 64 & 61 & 52 & 53 & 51 & 51 & 38 & 39 & 37 \\
4-- & 159 & 56 & 24 & 62 & 65 & \textbf{70} & 62 & 59 & 52 & 39 & 37 & 10 \\
\hline
\end{tabular}
}%end: setlength
\end{center}
\end{table}

\begin{table}[ht]
\caption{Instances solved in set
$\instancesprefilteredNEW$ preprocessed by \bloqqer with respect to classes by
number of qblocks (\#q) and number of formulas in each class (\#f).}
\label{fig:2017:exp:bloqqer37:prepro:solved:qblocks:full:cobra}
\begin{center}
{\setlength\tabcolsep{0.15cm}
\begin{tabular}{lrrrrrrrrrrrr}
\hline
\#q & \#f & \begin{turn}{80}\rareqs\end{turn}
& \begin{turn}{80}\caqe\end{turn}
& \begin{turn}{80}\heretic\end{turn} 
& \begin{turn}{80}\ijtihad\end{turn} & \begin{turn}{80}\revqfun\end{turn}
& \begin{turn}{80}\qstsdefnobreaksym\end{turn} 
& \begin{turn}{80}\qesto\end{turn} 
& \begin{turn}{80}\depqbfprefixopt\end{turn}
& \begin{turn}{80}\ghostqcegar\end{turn} & \begin{turn}{80}\quterandom\end{turn}
& \begin{turn}{80}\dynqbf\end{turn} \\
\hline
2 & 65 & 16 & 15 & 13 & 10 & 6 & 11 & 14 & 7 & 3 & 4 & \textbf{24} \\
3 & 218 & 80 & \textbf{81} & 65 & 59 & 65 & 46 & 50 & 34 & 53 & 23 & 18 \\
4--10 & 59 & 37 & 26 & \textbf{38} & 35 & 29 & 25 & 13 & 14 & 5 & 10 & 13 \\
11-20 & 53 & 25 & 25 & \textbf{31} & 17 & 22 & 24 & 18 & 30 & 9 & 21 & 4 \\
21-- & 42 & 17 & \textbf{22} & 17 & 15 & 13 & 21 & 20 & 17 & 12 & 15 & 6 \\
\hline
2--3 & 283 & \textbf{96} & \textbf{96} & 78 & 69 & 71 & 57 & 64 & 41 & 56 & 27 & 42 \\
4-- & 154 & 79 & 73 & \textbf{86} & 67 & 64 & 70 & 51 & 61 & 26 & 46 & 23 \\
\hline
\end{tabular}
}%end: setlength
\end{center}
\end{table}

\begin{table}[ht]
\caption{ Instances
solved in set $\instancesnoprefilteredHQSpreNEW$ filtered by \hqspre 
with respect to classes by number of qblocks (\#q) and number of formulas in
each class (\#f). }
\label{fig:2017:exp:hqspre:noprepro:solved:qblocks:full:cobra}
\begin{center}
{\setlength\tabcolsep{0.15cm}
\begin{tabular}{lrrrrrrrrrrrr}
\hline
\#q & \#f & 
\begin{turn}{80}\ghostqcegar\end{turn} & 
\begin{turn}{80}\revqfun\end{turn} & 
\begin{turn}{80}\caqe\end{turn} & 
\begin{turn}{80}\depqbfprefixopt\end{turn} & 
\begin{turn}{80}\qstsdefnobreaksym\end{turn} & 
\begin{turn}{80}\rareqs\end{turn} & 
\begin{turn}{80}\heretic\end{turn} & 
\begin{turn}{80}\quterandom\end{turn} & 
\begin{turn}{80}\dynqbf\end{turn} &
\begin{turn}{80}\qesto\end{turn} & 
\begin{turn}{80}\ijtihad\end{turn} \\ 
  \hline
2 & 70 & \textbf{36} & 18 & 5 & 7 & 9 & 2 & 2 & 2 & 20 & 4 & 2 \\
3 & 145 & 62 & \textbf{71} & 33 & 23 & 27 & 33 & 23 & 24 & 18 & 23 & 22 \\
4--10 & 26 & 3 & 5 & \textbf{7} & \textbf{7} & 1 & 4 & 5 & 5 & 4 & 4 & 3 \\
11--20 & 30 & 3 & 5 & 8 & \textbf{16} & 7 & 5 & 9 & 7 & 1 & 6 & 2 \\
21-- & 41 & 8 & 11 & \textbf{15} & 11 & 12 & 6 & 10 & 9 & 3 & 8 & 7 \\
\hline
2--3 & 215 & \textbf{98} & 89 & 38 & 30 & 36 & 35 & 25 & 26 & 38 & 27 & 24 \\
4-- & 97& 14 & 21 & 30 & \textbf{34} & 20 & 15 & 24 & 21 & 8 & 18 & 12 \\
\hline
\end{tabular}
}%end: setlength
\end{center}
\end{table}

\begin{table}[ht]
\caption{ Instances
solved in set $\instancesprefilteredHQSpreNEW$ preprocessed by \hqspre 
with respect to classes by number of qblocks (\#q) and number of formulas in
each class (\#f). }
\label{fig:2017:exp:hqspre:prepro:solved:qblocks:full:cobra}
\begin{center}
{\setlength\tabcolsep{0.15cm}
\begin{tabular}{lrrrrrrrrrrrr}
\hline
\#q & \#f & 
\begin{turn}{80}\caqe\end{turn} & 
\begin{turn}{80}\rareqs\end{turn} & 
\begin{turn}{80}\qesto\end{turn} & 
\begin{turn}{80}\revqfun\end{turn} & 
\begin{turn}{80}\heretic\end{turn} & 
\begin{turn}{80}\qstsdefnobreaksym\end{turn} & 
\begin{turn}{80}\depqbfprefixopt\end{turn} & 
\begin{turn}{80}\quterandom\end{turn} & 
\begin{turn}{80}\ijtihad\end{turn} &
\begin{turn}{80}\ghostqcegar\end{turn} & 
\begin{turn}{80}\dynqbf\end{turn} \\
\hline
2 & 70 & 18 & 15 & 15 & 14 & 15 & 17 & 15 & 15 & 11 & 7 & \textbf{24} \\
3 & 145 & \textbf{67} & 65 & 53 & 54 & 42 & 35 & 24 & 31 & 30 & 40 & 14 \\
4--10 & 26 & 6 & 6 & 8 & 4 & \textbf{10} & 2 & 7 & 5 & 5 & 1 & 5 \\
11--20 & 40 & 14 & 11 & 14 & 13 & 15 & 13 & \textbf{20} & 11 & 7 & 6 & 2 \\
21-- & 31 & \textbf{9} & 6 & 7 & 5 & 5 & 5 & 6 & 8 & 5 & 4 & 0 \\
\hline
2--3 & 215 & \textbf{85} & 80 & 68 & 68 & 57 & 52 & 39 & 46 & 41 & 47 & 38 \\
4-- & 97 & 29 & 23 & 29 & 22 & 30 & 20 & \textbf{33} & 24 & 17 & 11 & 7 \\
\hline
\end{tabular}
}%end: setlength
\end{center}
\end{table}

%%%%%%%%%%%%%%%%%%%%%%%%%%%%%%%%%%%%%%%%%%%%%%%%%%%%%%%%%%%%%%%%%%%%%%%

\clearpage

\subsection{Class-Based Analysis of Paradigms}

\begin{table}[ht]
\caption{Instances solved by solving
  paradigms in set
$\instancesnoprefilteredNEW$ filtered by \bloqqer with respect to classes by
number of qblocks (\#q) and number of formulas in each class (\#f).}
\label{fig:2017:exp:bloqqer37:noprepro:paradigms:solved:qblocks:full:cobra}
\begin{center}
{\setlength\tabcolsep{0.15cm}
\begin{tabular}{lrrrrrrr}
\hline
\#q & \#f & \ref*{paradigm:exp} & \ref*{paradigm:ghostq} & \ref*{paradigm:qsts} & \ref*{paradigm:caqe} & \ref*{paradigm:depqbf} & \ref*{paradigm:heretic} \\
\hline
2 & 63 & 26 & \textbf{32} & 8 & 6 & 7 & 2 \\
3 & 215 & \textbf{121} & 89 & 43 & 61 & 66 & 50 \\
4--10 & 63 & \textbf{38} & 4 & 21 & 27 & 16 & 34 \\
11--20 & 36 & 10 & 3 & 8 & 10 & \textbf{20} & 11 \\
21-- & 60 & \textbf{33} & 17 & 23 & 30 & 28 & 25 \\
\hline
2--3 & 278 & \textbf{147} & 121 & 51 & 67 & 73 & 52 \\
4-- & 159 & \textbf{81} & 24 & 52 & 67 & 64 & 70 \\
\hline
2-- & 437 & \textbf{228} & 145 & 103 & 134 & 137 & 122 \\
\hline
\end{tabular}
}%end: setlength
\end{center}
\end{table}

\begin{table}[ht]
\caption{Instances solved by solving
  paradigms in set
$\instancesprefilteredNEW$ preprocessed by \bloqqer with respect to classes by
number of qblocks (\#q) and number of formulas in each class (\#f).}
\label{fig:2017:exp:bloqqer37:prepro:paradigms:solved:qblocks:full:cobra}
\begin{center}
{\setlength\tabcolsep{0.15cm}
\begin{tabular}{lrrrrrrr}
\hline
\#q & \#f & \ref*{paradigm:exp} & \ref*{paradigm:ghostq} & \ref*{paradigm:qsts} & \ref*{paradigm:caqe} & \ref*{paradigm:depqbf} & \ref*{paradigm:heretic} \\
\hline
2 & 65 & \textbf{37} & 3 & 11 & 17 & 10 & 13 \\
3 & 218 & \textbf{103} & 53 & 46 & 86 & 40 & 65 \\
4--10 & 59 & \textbf{49} & 5 & 25 & 28 & 18 & 38 \\
11--20 & 53 & 31 & 9 & 24 & 27 & \textbf{32} & 31 \\
21-- & 42 & 18 & 12 & 21 & \textbf{24} & 17 & 17 \\
\hline
2--3 & 283 & \textbf{140} & 56 & 57 & 103 & 50 & 78 \\
4-- & 154 & \textbf{98} & 26 & 70 & 79 & 67 & 86 \\
\hline
2-- & 437 & \textbf{238} & 82 & 127 & 182 & 117 & 164 \\
\hline
\end{tabular}
}%end: setlength
\end{center}
\end{table}

\begin{table}[ht]
\caption{Instances solved by solving
  paradigms in set
$\instancesnoprefilteredHQSpreNEW$ filtered by \hqspre with respect to classes by
number of qblocks (\#q) and number of formulas in each class (\#f).}
\label{fig:2017:exp:hqspre:noprepro:paradigms:solved:qblocks:full:cobra}
\begin{center}
{\setlength\tabcolsep{0.15cm}
\begin{tabular}{lrrrrrrr}
\hline
\#q & \#f & \ref*{paradigm:exp} & \ref*{paradigm:ghostq} & \ref*{paradigm:qsts} & \ref*{paradigm:caqe} & \ref*{paradigm:depqbf} & \ref*{paradigm:heretic} \\
\hline
2 & 70 & 28 & \textbf{36} & 9 & 6 & 8 & 2 \\
3 & 145 & \textbf{85} & 62 & 27 & 36 & 40 & 23 \\
4--10 & 26 & \textbf{9} & 3 & 1 & \textbf{9} & 8 & 5 \\
11-20 & 30 & 8 & 3 & 7 & 8 & \textbf{16} & 9 \\
21-- & 41 & \textbf{15} & 8 & 12 & \textbf{15} & 11 & 10 \\
\hline
2--3 & 215 & \textbf{113} & 98 & 36 & 42 & 48 & 25 \\
4-- & 97 & 32 & 14 & 20 & 32 & \textbf{35} & 24 \\
\hline
2-- & 312 & \textbf{145} & 112 & 56 & 74 & 83 & 49 \\
\hline
\end{tabular}
}%end: setlength
\end{center}
\end{table}

\begin{table}[ht]
\caption{Instances solved by solving
  paradigms in set
$\instancesprefilteredHQSpreNEW$ preprocessed by \hqspre with respect to classes by
number of qblocks (\#q) and number of formulas in each class (\#f).}
\label{fig:2017:exp:hqspre:prepro:paradigms:solved:qblocks:full:cobra}
\begin{center}
{\setlength\tabcolsep{0.15cm}
\begin{tabular}{lrrrrrrr}
\hline
\#q & \#f & \ref*{paradigm:exp} & \ref*{paradigm:ghostq} & \ref*{paradigm:qsts} & \ref*{paradigm:caqe} & \ref*{paradigm:depqbf} & \ref*{paradigm:heretic} \\
\hline
2 & 70 & \textbf{37} & 7 & 17 & 18 & 21 & 15 \\
3 & 145 & \textbf{78} & 40 & 35 & 71 & 40 & 42 \\
4--10 & 26 & 10 & 1 & 2 & \textbf{13} & 7 & 10 \\
11--20 & 40 & 17 & 6 & 13 & 15 & \textbf{21} & 15 \\
21-- & 31 & 8 & 4 & 5 & \textbf{10} & 8 & 5 \\
\hline
2--3 & 215 & \textbf{115} & 47 & 52 & 89 & 61 & 57 \\
4-- & 97 & 35 & 11 & 20 & \textbf{38} & 36 & 30 \\
\hline
2-- & 312 & \textbf{150} & 58 & 72 & 127 & 97 & 87 \\
\hline
\end{tabular}
}%end: setlength
\end{center}
\end{table}

\clearpage
\subsection{VBS Statistics: Individual Solvers}

\begin{table}[ht]
\caption{Number of instances solved by the
  virtual best solver (VBS) in classes by number of qblocks (\#q), number of
  formulas (\#f) in each class, and relative contribution (percentage) of each solver to
  instances solved by the VBS for sets
  $\instancesnoprefilteredNEW$
  (\ref{fig:2017:exp:all:in:one:qblocks:vbs:cobra:noprepro:bloqqer:appendix}) 
  and $\instancesprefilteredNEW$~(\ref{fig:2017:exp:all:in:one:qblocks:vbs:cobra:prepro:bloqqer:appendix}).}
\addtocounter{table}{-1}
{\scriptsize
\subfloat[Set $\instancesnoprefilteredNEW$ filtered by \bloqqer.]{
{\setlength\tabcolsep{0.15cm}
\begin{tabular}{lrrrrrrrrrrrrr}
\hline
\#q & \#f & \begin{turn}{80}VBS\end{turn} & 
\begin{turn}{80}\revqfun\end{turn} & 
\begin{turn}{80}\ghostqcegar\end{turn} & 
\begin{turn}{80}\rareqs\end{turn} & 
\begin{turn}{80}\caqe\end{turn} & 
\begin{turn}{80}\heretic\end{turn} & 
\begin{turn}{80}\depqbfprefixopt\end{turn} & 
\begin{turn}{80}\ijtihad\end{turn} & 
\begin{turn}{80}\qstsdefnobreaksym\end{turn} & 
\begin{turn}{80}\quterandom\end{turn} & 
\begin{turn}{80}\qesto\end{turn} & 
\begin{turn}{80}\dynqbf\end{turn} \\
\hline
2 & 63 & 41 & 4.8 & \textbf{39.0} & 0.0 & 7.3 & 0.0 & 4.8 & 0.0 & 7.3 & 0.0 & 2.4 & 34.1  \\
3 & 215 & 133 & 26.3 & 5.2 & \textbf{30.0} & 6.0 & 1.5 & 10.5 & 0.7 & 12.7 & 2.2 & 3.0 & 1.5  \\
4--10 & 63 & 51 & 0.0 & 1.9 & 0.0 & 9.8 & 21.5 & 15.6 & \textbf{37.2} & 0.0 & 5.8 & 3.9 & 3.9  \\
11-20 & 36 & 23 & 0.0 & 0.0 & 4.3 & 8.6 & 0.0 & \textbf{56.5} & 0.0 & 26.0 & 0.0 & 0.0 & 4.3  \\
21-- & 60 & 40 & 22.5 & 2.5 & 15.0 & 7.5 & 2.5 & 10.0 & 0.0 & \textbf{25.0} & 10.0 & 0.0 & 5.0  \\
\hline
2--3 & 278 & 174 & 21.2 & 13.2 & \textbf{22.9} & 6.3 & 1.1 & 9.1 & 0.5 & 11.4 & 1.7 & 2.8 & 9.1  \\
4-- & 159 & 114 & 7.8 & 1.7 & 6.1 & 8.7 & 10.5 & \textbf{21.9} & 16.6 & 14.0 & 6.1 & 1.7 & 4.3  \\
\hline
2-- & 437 & 288 & 15.9 & 8.6 & \textbf{16.3} & 7.2 & 4.8 & 14.2 & 6.9 & 12.5 & 3.4 & 2.4 & 7.2  \\
\hline
\end{tabular}
}%end: setlength
\label{fig:2017:exp:all:in:one:qblocks:vbs:cobra:noprepro:bloqqer:appendix}
}%end subfloat

%%%%%%%%%

\subfloat[Set $\instancesprefilteredNEW$ preprocessed by \bloqqer.]{
{\setlength\tabcolsep{0.15cm}
\begin{tabular}{lrrrrrrrrrrrrr}
\hline
\#q & \#f & \begin{turn}{80}VBS\end{turn} & 
\begin{turn}{80}\rareqs\end{turn} & 
\begin{turn}{80}\caqe\end{turn} & 
\begin{turn}{80}\heretic\end{turn} & 
\begin{turn}{80}\ijtihad\end{turn} & 
\begin{turn}{80}\revqfun\end{turn} & 
\begin{turn}{80}\qstsdefnobreaksym\end{turn} & 
\begin{turn}{80}\qesto\end{turn} & 
\begin{turn}{80}\depqbfprefixopt\end{turn} & 
\begin{turn}{80}\ghostqcegar\end{turn} & 
\begin{turn}{80}\quterandom\end{turn} & 
\begin{turn}{80}\dynqbf\end{turn} \\
\hline
2 & 65 & 40 & 17.5 & 7.5 & 5.0 & 12.5 & 0.0 & 12.5 & 2.5 & 0.0 & 0.0 & 2.5 & \textbf{40.0}  \\
3 & 218 & 116 & \textbf{36.2} & 13.7 & 7.7 & 6.0 & 3.4 & 10.3 & 6.0 & 3.4 & 1.7 & 2.5 & 8.6  \\
4--10 & 59 & 54 & 0.0 & 0.0 & 27.7 & \textbf{29.6} & 7.4 & 3.7 & 3.7 & 16.6 & 0.0 & 0.0 & 11.1  \\
11--20 & 53 & 39 & \textbf{25.6} & 2.5 & 17.9 & 10.2 & 0.0 & 15.3 & 2.5 & 20.5 & 0.0 & 2.5 & 2.5  \\
21-- & 42 & 26 & 19.2 & 11.5 & 3.8 & 11.5 & 0.0 & \textbf{30.7} & 0.0 & 19.2 & 0.0 & 3.8 & 0.0  \\
\hline
2--3 & 283 & 156 & \textbf{31.4} & 12.1 & 7.0 & 7.6 & 2.5 & 10.8 & 5.1 & 2.5 & 1.2 & 2.5 & 16.6  \\
4-- & 154 & 119 & 12.6 & 3.3 & \textbf{19.3} & 19.3 & 3.3 & 13.4 & 2.5 & 18.4 & 0.0 & 1.6 & 5.8  \\
\hline
2-- & 437 & 275 & \textbf{23.2} & 8.3 & 12.3 & 12.7 & 2.9 & 12.0 & 4.0 & 9.4 & 0.7 & 2.1 & 12.0  \\
\hline
\end{tabular}
}%end: setlength
\label{fig:2017:exp:all:in:one:qblocks:vbs:cobra:prepro:bloqqer:appendix}
}%end subfloat

%%%%%%%%%

}%end:scriptsize
\label{fig:2017:exp:all:in:one:qblocks:vbs:cobra:appendix}
\refstepcounter{table}
\end{table}

\clearpage
\subsection{VBS Statistics: Paradigm-Based}

\begin{table}[ht]
  \caption{
    Number of instances solved by the
  virtual best solver (VBS) in classes by number of qblocks (\#q), number of
  formulas (\#f) in each class, and relative contribution (percentage) of each solving paradigm to
  instances solved by the VBS for set 
  $\instancesnoprefilteredNEW$ filtered by \bloqqer.
  }
\label{fig:2017:exp:bloqqer37:noprepro:paradigms:vbs:qblocks:full:cobra}
\begin{center}
{\setlength\tabcolsep{0.15cm}
\begin{tabular}{lrrrrrrrr}
\hline
\#q & \#f & VBS & \ref*{paradigm:exp} & \ref*{paradigm:ghostq} & \ref*{paradigm:qsts} & \ref*{paradigm:caqe} & \ref*{paradigm:depqbf} & \ref*{paradigm:heretic} \\
\hline
2 & 63 & 41 & \textbf{39.0} & \textbf{39.0} & 7.3 & 9.7 & 4.8 & 0.0  \\
3 & 215 & 133 & \textbf{58.6} & 5.2 & 12.7 & 9.0 & 12.7 & 1.5  \\
4--10 & 63 & 51 & \textbf{41.1} & 1.9 & 0.0 & 13.7 & 21.5 & 21.5  \\
11--20 & 36 & 23 & 8.6 & 0.0 & 26.0 & 8.6 & \textbf{56.5} & 0.0  \\
21-- & 60 & 40 & \textbf{42.5} & 2.5 & 25.0 & 7.5 & 20.0 & 2.5  \\
\hline
2--3 & 278 & 174 & \textbf{54.0} & 13.2 & 11.4 & 9.1 & 10.9 & 1.1  \\
4-- & 159 & 114 & \textbf{35.0} & 1.7 & 14.0 & 10.5 & 28.0 & 10.5  \\
\hline
2-- & 437 & 288 & \textbf{46.5} & 8.6 & 12.5 & 9.7 & 17.7 & 4.8  \\
\hline
\end{tabular}
}%end: setlength
\end{center}
\end{table}

\begin{table}[ht]
  \caption{
        Number of instances solved by the
  virtual best solver (VBS) in classes by number of qblocks (\#q), number of
  formulas (\#f) in each class, and relative contribution (percentage) of each solving paradigm to
  instances solved by the VBS for set 
  $\instancesprefilteredNEW$ preprocessed by \bloqqer.
}
\label{fig:2017:exp:bloqqer37:prepro:paradigms:vbs:qblocks:full:cobra}
\begin{center}
{\setlength\tabcolsep{0.15cm}
\begin{tabular}{lrrrrrrrr}
\hline
\#q & \#f & VBS & \ref*{paradigm:exp} & \ref*{paradigm:ghostq} & \ref*{paradigm:qsts} & \ref*{paradigm:caqe} & \ref*{paradigm:depqbf} & \ref*{paradigm:heretic} \\
\hline
2 & 65 & 40 & \textbf{70.0} & 0.0 & 12.5 & 10.0 & 2.5 & 5.0  \\
3 & 218 & 116 & \textbf{54.3} & 1.7 & 10.3 & 19.8 & 6.0 & 7.7  \\
4--10 & 59 & 54 & \textbf{48.1} & 0.0 & 3.7 & 3.7 & 16.6 & 27.7  \\
11--20 & 53 & 39 & \textbf{38.4} & 0.0 & 15.3 & 5.1 & 23.0 & 17.9  \\
21-- & 42 & 26 & \textbf{30.7} & 0.0 & \textbf{30.7} & 11.5 & 23.0 & 3.8  \\
\hline
2--3 & 283 & 156 & \textbf{58.3} & 1.2 & 10.8 & 17.3 & 5.1 & 7.0  \\
4-- & 154 & 119 & \textbf{41.1} & 0.0 & 13.4 & 5.8 & 20.1 & 19.3  \\
\hline
2-- & 437 & 275 & \textbf{50.9} & 0.7 & 12.0 & 12.3 & 11.6 & 12.3  \\
\hline
\end{tabular}
}%end: setlength
\end{center}
\end{table}

%%%%%%%%%%%%%%%%%%%%%%%%%%%%%%%%%%%%%%%%%%%%%%%%%%%%%%%%%%%%%%%%%%%%%%%%%%%%%%%%

\end{appendix}

\end{document}